\documentclass[conference,compsoc]{IEEEtran}
\IEEEoverridecommandlockouts

\usepackage{cite}
\usepackage{amsmath,amssymb,amsfonts}
\usepackage{algorithmic}
\usepackage{graphicx}
\usepackage{textcomp}
\usepackage{xcolor}
\def\BibTeX{{\rm B\kern-.05em{\sc i\kern-.025em b}\kern-.08em
    T\kern-.1667em\lower.7ex\hbox{E}\kern-.125emX}}

\PassOptionsToPackage{hyphens}{url}\usepackage{hyperref}
\usepackage[OT1]{fontenc} 
\usepackage{booktabs}
\usepackage{multirow}

\usepackage{listings}

\usepackage{pifont}
\usepackage{cancel}

\usepackage{amsfonts}
\usepackage{tikz}

\newcommand{\eg}{\textit{e.g.,} }

\usepackage{xcolor}
\usepackage{framed}
\definecolor{lightgray}{rgb}{0.95, 0.95, 0.95}
\definecolor{darkgray}{rgb}{0.6, 0.6, 0.6}

\usepackage{pgfplots}
\pgfplotsset{compat=1.17}
\usepgfplotslibrary{groupplots}
\PassOptionsToPackage{hyphens}{url}\usepackage{hyperref}

\usepackage{wasysym}
\newcommand{\ctrlFull}{\CIRCLE}
\newcommand{\ctrlHalf}{\LEFTcircle}
\newcommand{\ctrlOpen}{\Circle}

\definecolor{prLow}{rgb}{0.85,0.93,0.83}
\definecolor{prMod}{rgb}{1.00,0.91,0.74}
\definecolor{prHigh}{rgb}{0.96,0.78,0.76}
\definecolor{prKern}{rgb}{0.85,0.80,0.94}
\newcommand{\pvbadge}[2]{\colorbox{#1}{\small\textsf{#2}}}
\newcommand{\pvL}{\pvbadge{prLow}{L}}
\newcommand{\pvM}{\pvbadge{prMod}{M}}
\newcommand{\pvH}{\pvbadge{prHigh}{H}}
\newcommand{\pvK}{\pvbadge{prKern}{K}}

\definecolor{takeFrame}{rgb}{0.27,0.51,0.78}
\definecolor{takeFill}{rgb}{0.97,0.98,1.00}
\newenvironment{takeaway}{%
  \MakeFramed{\advance\hsize-\width\FrameRestore}%
  \noindent\hspace{-4pt}%
  \begin{list}{}{\setlength\leftmargin{0pt}\setlength\rightmargin{6pt}}%
  \item\relax\vspace{1pt}%
  \textcolor{takeFrame}{\textsc{Takeaway.}}\enskip\ignorespaces
}{%
  \vspace{1pt}\end{list}\endMakeFramed%
}

\newcommand{\numSampleReport}{100\xspace}

\newcommand{\chromeTotal}{2,770\xspace}

\newcommand{\datasetMC}{2,233\xspace}

\usepackage{colortbl}

\definecolor{hmZero}{rgb}{1.00,1.00,1.00}
\definecolor{hmOne}{rgb}{0.93,0.95,0.99}
\definecolor{hmTwo}{rgb}{0.83,0.89,0.97}
\definecolor{hmThree}{rgb}{0.66,0.80,0.93}
\definecolor{hmFour}{rgb}{0.45,0.66,0.86}
\definecolor{hmFive}{rgb}{0.27,0.51,0.78}
\newcommand{\hZ}[1]{\cellcolor{hmZero}#1}
\newcommand{\hA}[1]{\cellcolor{hmOne}#1}
\newcommand{\hB}[1]{\cellcolor{hmTwo}#1}
\newcommand{\hC}[1]{\cellcolor{hmThree}#1}
\newcommand{\hD}[1]{\cellcolor{hmFour}#1}
\newcommand{\hE}[1]{\textcolor{white}{\cellcolor{hmFive}\textbf{#1}}}

\definecolor{clsBinary}{HTML}{4E9C81}
\definecolor{clsDoc}{HTML}{3A6EA5}
\definecolor{clsScript}{HTML}{E08E45}
\definecolor{clsUI}{HTML}{8E7CC3}
\definecolor{clsIPC}{HTML}{C0504D}

\usepackage{xspace}

\newcommand{\totalreports}{4099\xspace}

\newcommand{\firefoxreports}{1329\xspace}

\usepackage{fontawesome5}

\begin{document}

\title{
SoK: Taxonomizing the Low-Level Attack Surface of Modern Web Browsers  
}

\author{\IEEEauthorblockN{Han Zheng}
\IEEEauthorblockA{%
\textit{EPFL}\\
}
\and 
\IEEEauthorblockN{Qinying Wang}
\IEEEauthorblockA{%
\textit{EPFL}\\
}
\and
\IEEEauthorblockN{Qiang Liu}
\IEEEauthorblockA{%
\textit{EPFL}\\
}
\and 
\IEEEauthorblockN{Mathias Payer}
\IEEEauthorblockA{%
\textit{EPFL}\\
}
}

\maketitle

\sloppy

\begin{abstract}

The web browser remains one of the most exposed remote attack
surfaces on end-user systems, and memory-corruption flaws continue
to play a central role in real-world browser exploitation.
Despite a decade of intensive browser testing and 
bug-disclosure efforts, the community still lacks an explicit, 
defense-oriented systematization of the 
browser's low-level attack surface.
Prior SoKs have surveyed browser vulnerabilities and mitigation techniques. However, these
perspectives remain fragmented, leaving open a central question:
\emph{how is the low-level attack surface of modern web
browsers structured, and which parts of this surface remain
underexplored by existing security testing?}

We approach this primary question through three sub-questions.
(RQ1) How is the browser's attack surface structured
along input classes and components? (RQ2) Where do memory corruption 
vulnerabilities arise within this taxonomy? (RQ3) What do these 
attack-surface patterns imply for existing browser security testing? 
To answer RQ1, we derive an architecture-grounded
Input~$\times$~Component~$\times$~Privilege taxonomy that
abstracts the architectures of Chrome, Firefox, and Safari into
a unified view. To answer RQ2, we map \datasetMC
memory corruption reports disclosed between 2016 and 2025 
onto this taxonomy. To answer RQ3, we overlay
a decade of academic browser fuzzers, classified by the
targeted input class, onto the bug-density map.
Our systematization reveals that current
testing concentrates on well-explored components while bug-dense,
high-impact surfaces remain insufficiently tested. Moreover, we 
identify three fuzzer deployment gaps, which are orthogonal to 
the academic efforts. 
Our work offers a structured foundation for future browser security
research, so that researchers can use this systematization as (i)
a map of which inputs reach which components at which privilege,
(ii) a measurement of where a decade of bugs and testing efforts
concentrate, and (iii) a prioritization guide for where testing is
insufficient.

\end{abstract}

\section{Introduction}
The web browser is one of the most high-value remote attack target for
end-user devices. Memory corruption in the browser codebase is the
dominant exploit class, contributing 67\% of in-the-wild
exploit chains observed against consumer
devices~\cite{p0_itw_2021, googletag_state_back,
googletag_buying_spying}. 
The dominance of memory corruption motivates the focus on the
browser's \emph{low-level attack surface}: the native (C/C++)
code running in the browser processes that parses 
attacker-controlled bytes (such as HTML, JavaScript, images, IPC
messages, and UI events).
Each path from an attacker-controlled input into the code that
parses it forms one \emph{surface}, and the attack surface is
the union of these paths (\autoref{sec:scope-method}).
To test this surface,
browser vendors and academia have
responded with a decade of continuous fuzzing of both the browser
and the third-party libraries it depends
on~\cite{ossfuzz_introspector, clusterfuzz}, disclosing tens of
thousands of memory-safety bugs~\cite{chrome_vrp_2023, chrome_vrp_2024,
chrome_vrp_2025}.
However, these efforts target individual components in isolation,
and the resulting knowledge stays scattered across bug trackers,
fuzzer harnesses, and design documents.
The community therefore lacks (i) a systematic view of how the 
low-level attack surface is structured, and (ii) a defense-oriented view 
of how far existing testing efforts reach into this surface and 
where future testing should focus. This motivates our overarching research
question: \emph{how is the low-level attack surface of modern web
browsers structured, and which parts of this surface remain
underexplored by existing security testing?}

Prior SoK study browser security through different lenses. 
Lim et al.~\cite{lim2021sok} provide a broad view of browser bugs,
exploitation techniques, and mitigations. 
These perspectives organize browser security around 
bug primitives and mitigations. 
However, their systematization lacks a unified view of 
the low-level attack surface, \eg what are the browser inputs, 
which browser components process these inputs, and what are 
their privilege. Without this view, it is difficult to assess where memory-corruption risks 
concentrate, which surfaces are exercised by current testing 
techniques, and where future testing and hardening should be 
prioritized.
To close this gap, we split our overarching research question into 
three sub-questions.

\begin{figure*}[h!]
	\centering
  \includegraphics[width=\linewidth]{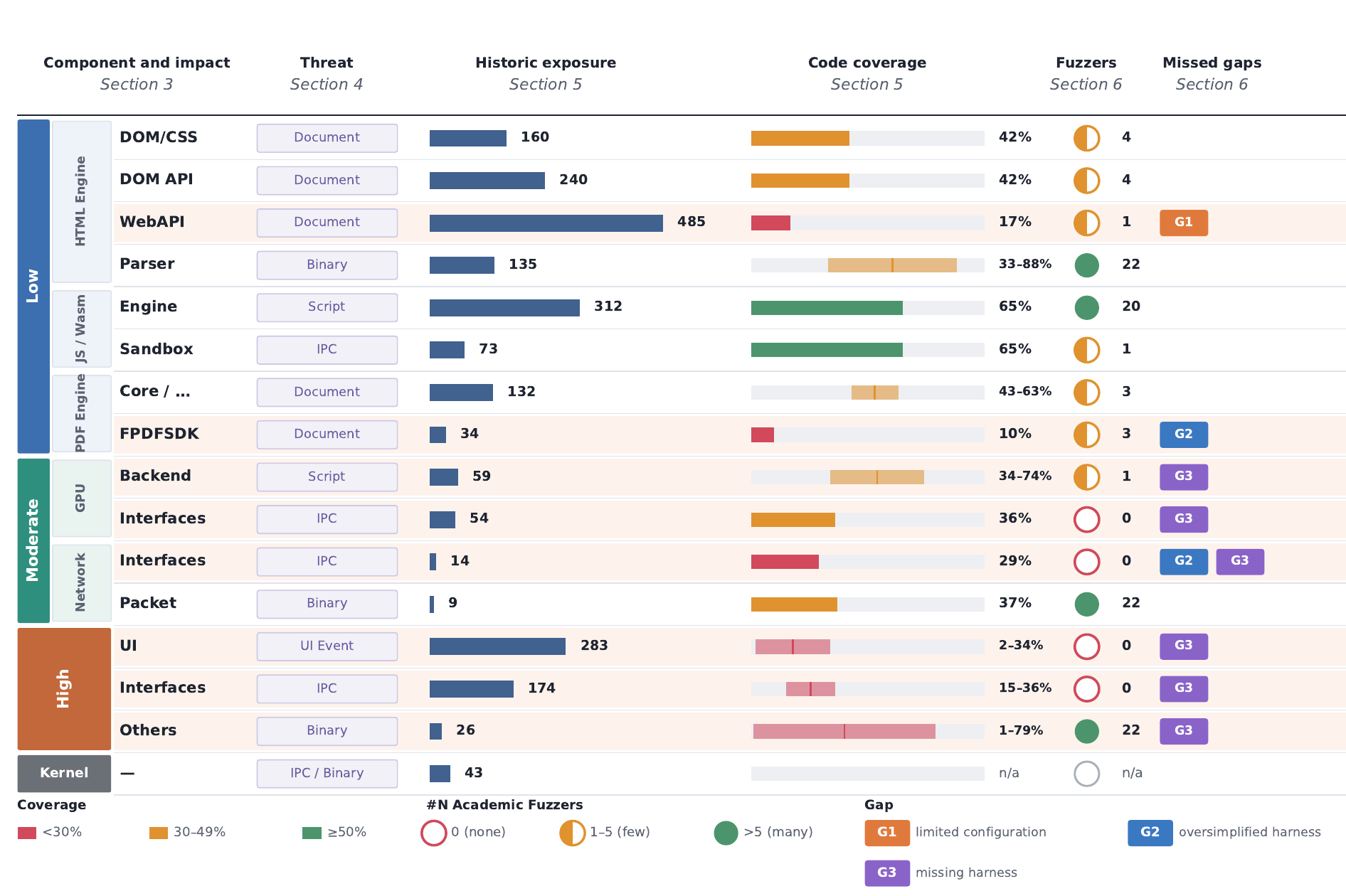}
  \caption{Systematization of the browser attack surface. 
  }
  \label{fig:method-systematization}
\end{figure*}

\textbf{RQ1: How is the browser's attack surface
structured along input classes and components?}
We systematize the browser's low-level attack
surface along an Input~$\times$~Component~$\times$~Privilege
taxonomy. Modern browsers follow the principle of least
privilege~\cite{chrome_least_privileg}, isolating the renderer,
GPU, network, and browser code into separate processes at
different OS privileges. The taxonomy mirrors this 
structure: it studies each component, its attacker-controlled
inputs, and its privilege tier in isolation across Chrome, 
Firefox, and Safari, rather than treating the browser as one monolith.
We derive it from the public design documents of all three browsers~\cite{chrome_proc_type, firefox_proc_type, safari_proc_type, chrome_least_privileg}.
The resulting taxonomy distills the three browsers into five
attacker-controlled input classes (binary blobs, documents,
scripts, UI gestures, and IPC calls), processed by nine
components across four privilege tiers.

\textbf{RQ2: Where do memory corruption vulnerabilities 
arise within this taxonomy?}
We answer RQ2 by collecting \datasetMC memory corruption reports
disclosed between 2016 and 2025 from the Chromium issue tracker
and Firefox security advisories~\cite{chrome_issue_tracker,
firefox_security_bulletin}, and classifying each against the
Input~$\times$~Component~$\times$~Privilege matrix.
We also pair every component with the line coverage that
vendor-deployed fuzzers achieve~\cite{chromium_cov_dashboard,ossfuzz_introspector}.
Five bug-dense surfaces drive the divergence between vendor testing
investment and bug count: the WebAPI bindings, the PDFium SDK
boundary, the WebGL backend, the UI input path, and the IPC receivers.
Together they hold a large share of high-privilege bugs yet
remain insufficiently tested.

\textbf{RQ3: What do these attack-surface patterns
imply for existing browser security testing?}
We overlay a decade of academic browser fuzzers, classified by
their targeted input class, onto the bug-density map from RQ2.
The overlay shows that academic fuzzers cluster on the script
and document inputs, so the bug-dense surfaces identified in
RQ2 stay under-covered. Beyond the testing techniques
themselves, we diagnose three recurring deployment gaps that
explain part of the shortfall: incorrect configuration,
oversimplified harnesses, and missing harnesses.

Overall, this paper presents the following major findings:

\begin{itemize}
  \item An architecture-grounded systematization of the
        browser's low-level attack surface: an
        Input~$\times$~Component~$\times$~Privilege taxonomy
        abstracted from Chrome, Firefox, and Safari.
  \item An empirical study of historical browser
        memory corruption vulnerabilities: \datasetMC reports
        from Chrome and Firefox (2016--2025) classified onto the
        taxonomy to locate bug-dense input-to-component paths.
  \item A defense-oriented systematization of well-tested
        and overlooked attack surfaces: vendor coverage and a
        decade of academic fuzzers overlaid on the bug-density
        map, exposing under-covered surfaces and three
        recurring deployment gaps.
  \item Future directions for browser security research:
        tighter Principle-of-Least-Privilege enforcement,
        expanded coverage of insufficiently tested attack
        surfaces, and low-cost deployment fixes.
\end{itemize}

\definecolor{prVL}{rgb}{0.78,0.88,0.78}
\newcommand{\pvVL}{\pvbadge{prVL}{VL}}

\newcommand{\scL}{\pvbadge{prLow}{Low}}
\newcommand{\scM}{\pvbadge{prMod}{Mod}}
\newcommand{\scH}{\pvbadge{prHigh}{High}}

\begin{table*}[t]
\centering
\caption{Components and interfaces of modern browsers. 
We define the component privilege following the vendor practice~\cite{chrome_severity_guideline}.
\textbf{Privilege:} \pvVL{}~very low, \pvL{}~low, \pvM{}~moderate,
\pvH{}~high.
\textbf{Scale:} We measure on Chrome codebase
(\scL{}~$<1$M, \scM{} $1$--$10$M LoC, \scH{}~$\geq 10$M LoC).
}
\label{tab:bg-component}
\small
\setlength{\tabcolsep}{4pt}
\resizebox{\linewidth}{!}{%
\begin{tabular}{l l l c c c c c}
\toprule
& & & \multicolumn{3}{c}{\textbf{Privilege}}
& \multirow{2}{*}{\textbf{Scale}}
& \multirow{2}{*}{\textbf{Process content}} \\
\cmidrule(lr){4-6}
\textbf{Impact} & \textbf{Component} & \textbf{Function}
 & Chrome & Firefox & Safari & & \\
\midrule
\multirow{5}{*}{\textsf{Low}}
& HTML engine    & HTML / CSS parsing, DOM, layout, paint
                 & \pvL & \pvL & \pvL
                 & \scH
                 & HTML Document \\
& PDF engine     & Parse and render PDF documents
                 & \pvL & \pvL & \pvL
                 & \scL
                 & PDF document \\
& JS engine      & JS / Wasm execution (in sandbox)
                 & \pvVL & \pvVL & \pvVL
                 & \scM
                 & Script \\
& Media decoder  & Decode images, audio, and video
                 & \pvL & \pvVL & \pvM
                 & \scM
                 & Binary blob \\
& Third-party parser & Parse fonts, XML, and other formats
                 & \pvL & \pvL & \pvL
                 & \scM
                 & Binary blob \\
\midrule
\multirow{3}{*}{\textsf{Moderate}}
& GPU            & WebGPU/WebGL backend + shader translation
                 & \pvM & \pvH$^{+}$ & \pvM
                 & \scH
                 & Shader script, IPC \\
& Network        & TCP/UDP + TLS + cert validation
                 & \pvM & \pvM & \pvM
                 & \scM
                 & Binary (packet, cert) \\
& Utility        & Privileged helper tasks (print, etc.)
                 & \pvM & \pvM & \pvH
                 & \scL
                 & IPC \\
\midrule
\textsf{High}
& Browser proc.  & Profile, navigation, Secure UI
                 & \pvH & \pvH & \pvH
                 & \scH
                 & UI gesture, IPC \\
\bottomrule
\end{tabular}%
}

\medskip
\footnotesize
$^{+}$~Firefox runs the GPU process unsandboxed except Windows.
\end{table*}

\section{Scope and Threat Model}
\label{sec:scope-method}

\textbf{Study scope.}
This paper examines the low-level attack surface of the web
browser and the corresponding testing techniques.
We define the attack surface as the union of paths from each
attacker-controlled input class into the native code region
that parses it. Each path forms one \emph{surface}, exercised
at that code's privilege tier, and \autoref{sec:threat-model}
enumerates these surfaces.
For the attack surface,
we study the three major web browsers: Chrome, Firefox, and Safari.
For the bug study, we explore memory corruption bugs, which are
the dominant source of in-the-wild exploits~\cite{p0_itw_2021}.
For the testing technique, we primarily study the fuzzing
technique and discuss the rest in \autoref{sec:discuss}.
\autoref{fig:method-systematization} summarizes our systematization.

\textbf{Threat model.}
Following vendor practices~\cite{chrome_severity_guideline}, 
we consider a remote attacker who tricks a user into clicking 
attacker-controlled web pages and into performing specific UI 
interactions. The attacker's goal is to corrupt the victim's 
web browser process memory, achieve code execution in one process, 
elevate privileges and ultimately achieve code execution 
in the browser process or even in the kernel. 
We also account for scenarios in which an attacker already has code 
execution capability inside a compromised low privilege process 
and sends crafted IPC commands to corrupt higher privilege 
processes~\cite{chrome_compromised_renderer}.

\begin{figure}[t!]
	\centering
  \includegraphics[width=\columnwidth]{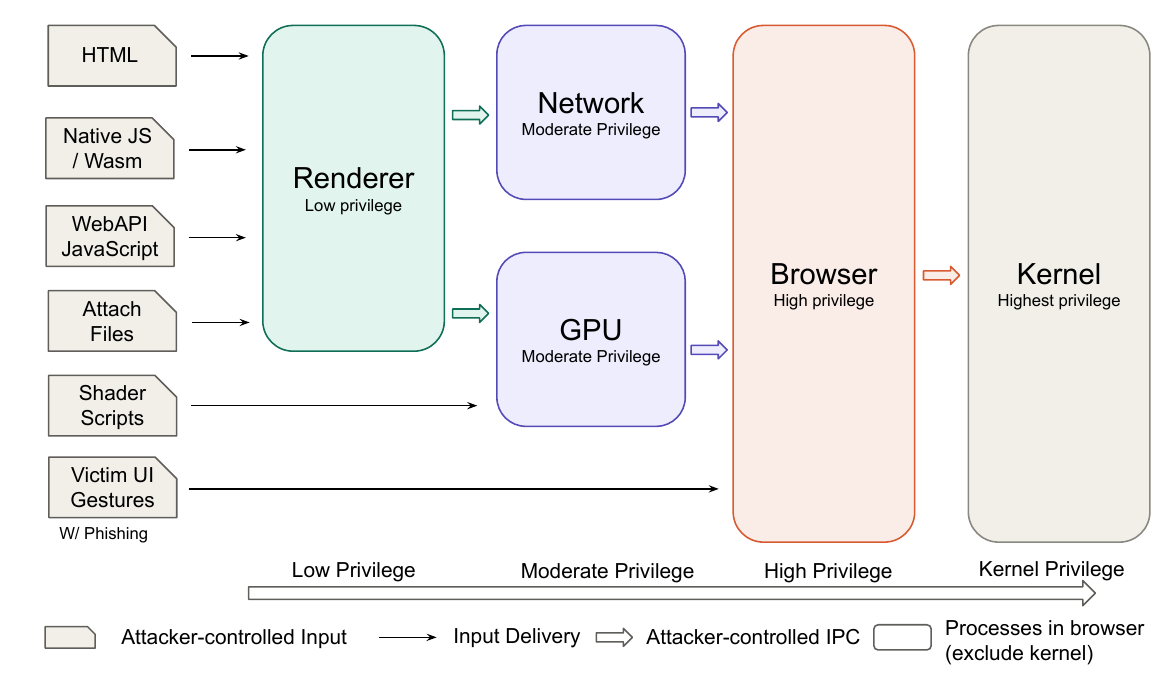}
  \caption{The browser attack surface. We only enumerate the main
  surfaces and processes. For instance, the network
  process may receive untrusted packets (binary blobs) from an
  attacker-controlled website.}
  \label{fig:input-flow}
\end{figure}

\section{Browser Components and Interfaces}
\label{sec:component}

To answer RQ1, we first examine the components and interfaces 
in the web browsers.
Modern browsers follow the principle of least privilege, running
each functional component in a separate process at its own
OS-level privilege tier, so that a memory bug in one process
cannot, by itself, corrupt another~\cite{chrome_least_privileg,
chrome_multi_proc}. We systematize how Chrome, Firefox, and
Safari enforce this partitioning by leveraging their design
documents~\cite{chrome_proc_type, firefox_proc_type,
safari_proc_type, chrome_least_privileg, chrome_multi_proc}.
The architecture includes two parts: components and interfaces. 
A \emph{component} is a subsystem that implements a specific browser 
function and processes a distinct class of attacker-controlled input. 
Typical browser components include 
the HTML engine, PDF engine, JavaScript engine,
media decoder, third-party parsers, GPU, network, utility, and
browser process, each at a defined privilege tier. 
Components communicate through inter-process communication (IPC), 
which we treat as an attacker-controlled channel under the 
compromised-renderer threat model~\cite{chrome_compromised_renderer}.
\autoref{tab:bg-component} summarizes each
component's role, scale, content, and privilege across the three
vendors. \autoref{fig:input-flow} illustrates how inputs flow
through them.

\textbf{HTML engine.}
The HTML engine parses HTML and CSS, builds the DOM tree, and
drives layout, paint, and the document API or WebAPI. 
All three browsers host it in a low-privilege renderer
process~\cite{chrome_sandbox, firefox_sandbox}. 
When the engine needs a privileged resource
(a file dialog, a network request, GPU acceleration), it requests
one from a higher-privilege component rather than acting
on the resource itself. Because it parses untrusted documents
directly from the network, the HTML engine is directly exposed 
to attacker-controlled input, and at roughly 10M lines of code, which is one of the largest single-purpose components in the browser.

\textbf{PDF engine.}
The PDF engine parses and renders embedded PDF documents. Chrome
and Safari ship dedicated native engines (PDFium in Chrome,
PDFKit in Safari) and host them inside the low-privilege
process~\cite{chrome_sandbox, safari_proc_type}. Firefox
takes a different route and renders PDFs via
PDF.js~\cite{firefox_pdfjs}, a JavaScript application that runs
inside the JavaScript engine itself. PDF documents are fully
attacker-shaped on all three browsers, so the PDF engine is
directly exposed to attacker-controlled bytes. At roughly 500K
lines of code it is only an small component in the browser.

\textbf{JavaScript engine.}
The JavaScript engine (V8 in Chrome, SpiderMonkey in Firefox,
JavaScriptCore in Safari~\cite{chrome_v8,safari_jsc,firefox_sm}) 
executes JavaScript and WebAssembly drawn from the loaded document. 
To bound the impact of the vulnerabilities, 
JavaScript code runs inside an internal
sandbox~\cite{chrome_v8_sandbox} that confines it to a
restricted memory region. This internal sandbox itself lives
inside the renderer process and shares its OS-level
low-privilege. The engine is fully exposed to
attacker-controlled bytes, and spans roughly 3M lines of code.

\textbf{Media decoder.}
The media decoder demuxes and decodes the binary image, audio,
and video formats that pages reference or embed (\eg PNG, JPEG,
WebP, AVIF, MP4, WebM). Chrome integrates image decoding
into the Tab process and offloads hardware-accelerated video
decoding to the GPU process when the platform supports
it~\cite{chrome_video_hardware_gpu}. Firefox runs a dedicated
RDD (Remote Data Decoder) process for video and a GMPPlugin
process for images at very low privilege~\cite{firefox_proc_design}.
Safari folds both into the GPU process at moderate privilege~\cite{safari_image_gpu,safari_media_gpu} to improve performance. 
The media decoder weighs in at roughly 4M lines of code, the bulk
of which sits in third-party codec and real-time communication 
libraries. 

\textbf{Third-party parser.}
Beyond the core decoders, the browser embeds
a long tail of third-party libraries that parse other
attacker-supplied formats: FreeType and HarfBuzz for fonts,
libxml and expat for XML, and similar parsers for additional
formats. All of them live inside the renderer sandbox at low
privilege. These parsers see fully attacker-controlled bytes and
have a long history of memory corruption disclosures. 
Together they add roughly 1M lines of code to the renderer.

\textbf{GPU.}
The GPU process drives the system's graphics driver, compiles
attacker-supplied shader programs, and executes
the resulting draw calls~\cite{chrome_proc_type,
safari_proc_type, firefox_proc_type}. Because it must call into
the closed-source GPU driver, the GPU sandbox has more permission 
than the renderer. On Firefox, the GPU process runs entirely
unsandboxed on non-Windows OSes~\cite{firefox_proc_design,
chrome_gpu_limited_sbx, chrome_sbx_by_platform}. The GPU
component is by far the largest in the browser at roughly 33M
lines of code, the bulk of which is WebGL and WebGPU backends. 

\textbf{Network.}
The network component owns the TCP stack, SSL/TLS handshake, and
certificate validation. All browsers, including Chrome, Firefox, and 
Safari isolate it into a dedicated network process at moderate
privilege~\cite{firefox_proc_design, safari_proc_type}.
One exception is Firefox, whose NSS cryptographic stack validates
certificates in the high-privilege parent
process~\cite{firefox_nss_bug}.
Although the network process mostly handles IPC, it receives
raw bytes from remote servers, which a malicious server can
fully control. The networking stack spans roughly 4.5M lines of
code.%

\textbf{Utility.}
Utility processes host functionality that needs more privilege
than the renderer but less than the browser~\cite{chrome_print_util}. 
Chrome and Firefox both run multiple utility processes at moderate 
privilege. Safari instead folds some tasks into the central UIProcess at high privilege. 
The utility tier accounts for roughly 280K lines of code, the
smallest among the components.

\textbf{Browser process.}
The browser process is the coordinator: it owns the
user-profile database, the cookie store, and the navigation bar,
and displays the results to the user. It also receives
direct user input events (mouse, keyboard, touch), which become
a partially attacker-controlled input under phishing or
social-engineering scenarios~\cite{chrome_severity_guideline}.
All three browsers run at the highest privilege
in the browser ecosystem~\footnote{The kernel is outside the
ecosystem, which we further discuss in~\autoref{sec:threat-model}.}. Counting the reusable browser modules it
hosts (autofill, password manager, safe browsing, and similar),
the browser process spans roughly 12M lines of code.

\textbf{Interfaces (IPC).}
Components in different processes communicate through the IPC: 
Chrome uses Mojo, Firefox uses IPDL, and Safari uses XPC. 
By design, every IPC message from a lower-privilege component to a
higher-privilege one must be validated by the receiver. In
practice, IPC handlers are a recurring source of memory corruption
bugs, because they expose privileged components to messages
crafted by an attacker who has already compromised the renderer
(the canonical sandbox-escape vector). We therefore treat IPC as
a first-class attacker-controlled input alongside web content,
not as a transport detail (\autoref{sec:threat-model}).

\begin{takeaway}
Modern browsers enforce the principle of least privilege by 
isolating the HTML, PDF, and JavaScript engines, media and
resource parsers, GPU, network, utility, and browser code 
into separate processes at different OS-level privileges.

\end{takeaway}

\begin{figure}[!t]
\centering
\includegraphics[width=\columnwidth]{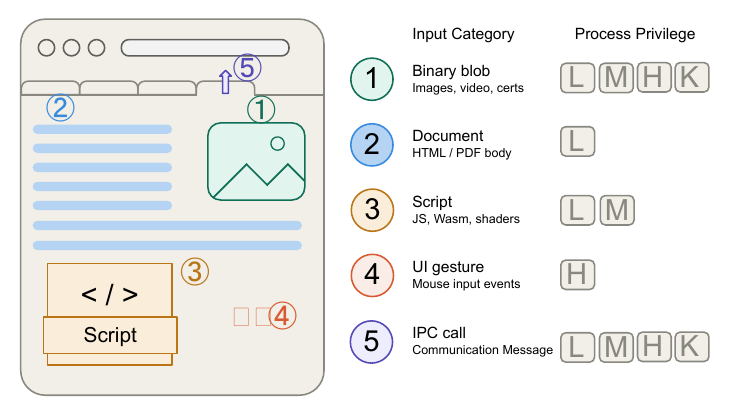}
\caption{The five categories of input a web browser accepts. The
arrow at the top denotes IPC requests issued by the renderer (Tab)
process; these become fully attacker-controlled once the attacker
achieves renderer code execution.}
\label{fig:tm-inputs}
\end{figure}

\section{Browser Attack Surface}
\label{sec:threat-model}

To complete the answer to RQ1, we build on the component and
privilege structure defined in~\autoref{sec:component} and
characterize the threats from the attacker's perspective: the
classes of input an attacker can control, and, for each class, the
component the input reaches and the privilege it thereby places at
risk, following the threat model in \autoref{sec:scope-method}.

We group the browser inputs into five classes, illustrated
in \autoref{fig:tm-inputs} and summarised in
\autoref{tab:tm-inputs}. Note that the table only discusses
the main components.
We further enumerate each class's surfaces: the inputs and
the components that parse them.

\begin{table}[t]
\centering
\caption{The five input classes, with representative formats, the
component primarily reached, the privilege exercised there, and the
degree of attacker control. 
Privilege is colour-coded:
\pvL{}~low, \pvM{}~moderate, \pvH{}~high and \pvK{}~kernel. 
Attacker control uses
distinct markers: \ctrlFull{}~full control; \ctrlHalf{}~full control
only once the sending process is compromised; \ctrlOpen{}~limited
control, requiring user participation.
*: HTML bodies are processed by the renderer process, but its 
document API and WebAPI calls can impact moderate/high privilege 
processes.}
\label{tab:tm-inputs}

\resizebox{\columnwidth}{!}{%
\begin{tabular}{l l l c c}
\toprule
Input class & Repre. Format & Target & Priv.\ & Control \\
\midrule
\multirow{3}{*}{Binary blob}
            & Image, font, Flash  & Renderer        & \pvL          & \ctrlFull \\
            & Video                 & Renderer, GPU             & \pvL\,\pvM    & \ctrlFull \\
            & Network packet        & Network, Kernel & \pvM\,\pvK    & \ctrlFull \\
            & Crypto Cert        & Network, Browser & \pvM\,\pvH    & \ctrlFull \\
\addlinespace
Document*    & HTML, PDF             & Renderer        & \pvL          & \ctrlFull \\
\addlinespace
\multirow{2}{*}{Script}
            & JavaScript            & Renderer        & \pvL          & \ctrlFull \\
            & Shader                & GPU             & \pvM          & \ctrlFull \\
\addlinespace
UI gesture  & Input event           & Browser         & \pvH          & \ctrlOpen \\
\addlinespace
\multirow{3}{*}{IPC call}
            & Sandbox IPC               & Renderer        & \pvL          & \ctrlHalf \\
            & Process IPC                   & Non-renderer    & \pvM\,\pvH    & \ctrlHalf \\
            & Syscall               & Kernel          & \pvK          & \ctrlHalf \\
\bottomrule
\end{tabular}%
}

\end{table}

\subsection{Attack Surface: Binary-Blob Inputs}
\label{ssec:tm-binary}

Web content carries files in diverse formats, including images,
fonts, database files, and video~\cite{chrome_libpng_bug,
chrome_freetype_bug,chrome_sqlite_bug,chrome_video_bug}. These inputs
reach the browser along two paths: embedded within the HTML document,
or delivered as a network payload from an attacker-controlled
website~\cite{firefox_net_bug,firefox_nss_bug}. Although each format
has its own specification, none imposes the strong structural
requirements of a full document format, so we group them together as
\emph{binary blobs}.

Because binary blobs arrive directly from untrusted websites, browsers
process most of them inside strictly sandboxed, low-privilege
processes, typically the renderer, to bound the impact of a
parsing-library vulnerability. Some blobs, however, are passed
directly to moderate- or high-privilege processes. A
hardware-accelerated video stream, for example, is demuxed in the
renderer but decoded in the GPU process, which drives the hardware
accelerator~\cite{chrome_video_hardware_gpu}.
Likewise, network packets and cryptographic certificates are processed
in the moderate-privilege network service or even high-privilege 
browser process rather than in the renderer.

\begin{takeaway}
Binary blobs primarily threaten low-privilege components, but certain
formats, such as hardware-accelerated video, network packets, and
cryptographic certificates, are handled directly in moderate- or
high-privilege processes.
\end{takeaway}

\subsection{Attack Surface: Document Inputs}
\label{ssec:tm-doc}

Both HTML and PDF documents consist of a body and embedded JavaScript,
and all three browsers treat the document as untrusted input.
The HTML and PDF engines (\autoref{sec:component}) parse the
body inside low-privilege processes in accordance with the
principle of least privilege~\cite{chrome_least_privileg}.

The threat surface of a document is not confined to its parser,
however, since embedded JavaScript can invoke APIs that call into
external modules. Here the two formats diverge sharply. HTML scripts
have access to both document APIs and WebAPI. Both can reach
external modules that run in higher-privilege processes. A document
that exercises the WebGPU or WebGL API, for instance, drives graphics
code hosted in the moderate-privilege GPU process, so attacker-supplied
content can reach and corrupt the moderate-privilege GPU process. 
PDF, by contrast, exposes a far
more restricted JavaScript API that operates only on the document
itself. Its scripts cannot reach beyond the renderer, so PDF content
remains confined to the low-privilege process.

\begin{takeaway}
Document bodies are parsed in low-privilege processes across all
vendors. However, HTML may include WebAPI and document API calls 
that propagate to higher-privilege processes.
\end{takeaway}

\subsection{Attack Surface: Script Inputs}
\label{ssec:tm-script}

Scripts are embedded in web content to enable web applications and
programmable graphics, and are therefore directly under attacker
control. Native JavaScript and WebAssembly (Wasm) 
are delivered straight to the JavaScript engine, which in all modern browsers runs inside a sandboxed,
low-privilege process. This placement confines a highly flexible
attacker input at an acceptable performance cost, and it reflects the
fact that JavaScript engines have historically been a rich source of
memory corruption bugs.

JavaScript additionally exposes WebAPI and document APIs, through
which a script can interact with other browser subsystems and target
higher-privilege components. We account for these paths under other
input classes: a WebAPI call that requires no special privilege is
folded into the document class (\autoref{ssec:tm-doc}), while a Mojo
WebAPI that presupposes a compromised renderer is classified as an IPC
call (\autoref{ssec:tm-ipc}). This subsection therefore concerns only
native JavaScript.

Beyond JavaScript, browsers also accept native graphics-shader
programs as script input. These are passed directly to the shader
translator hosted in the GPU process, which translates the shader into
a backend-specific language (\eg HLSL on Windows). The result is then
compiled by a vendor-provided compiler (\eg DXC on Windows) before
execution on the GPU. This pipeline allows attacker-controlled input
to reach a moderate-privilege process. We do not consider 
vendor-provided compilers in this work, as they are maintained by the
operating-system vendor and are in some cases closed-source~\cite{metal_close_source}.

\begin{takeaway}
Native JavaScript only threatens the sandboxed JavaScript
engine, whereas graphics shader programs
can carry attacker-controlled input into moderate-privilege
components such as the GPU process.
\end{takeaway}

\subsection{Attack Surface: UI Gestures}
\label{ssec:tm-ui}

UI gestures originate from the user (victim) and are, by 
default, treated as trusted input. They are mostly processed directly by 
the browser process, the most privileged component in the architecture. 
Vendor threat models nonetheless acknowledge that an attacker may 
persuade a user to perform a specific sequence of UI actions, 
through a phishing
page or other social engineering, so the input is partially, though
not fully, attacker-controlled~\cite{chrome_severity_guideline}. 
Unlike inputs that an attacker
controls directly, a UI-driven attack typically requires user
participation and a specific interaction sequence. When the required
sequence is sufficiently complex or implausible, vendors classify the
issue as a functional bug rather than a security vulnerability, which
bounds the practical reach of this class even though it targets the
high-privilege component.

\begin{takeaway}
UI gestures are only indirectly attacker-controlled, yet they are
processed by the high-privilege browser process.
\end{takeaway}

\subsection{Attack Surface: IPC Calls}
\label{ssec:tm-ipc}

Lower-privilege processes or components must request resources from
higher-privilege ones, and IPC is the mechanism for doing so. By
design, any IPC message originating from a lower-privilege process
should be treated as attacker-controlled and validated accordingly. In practice
this assumption is not always applied consistently, and an
insufficiently-validated IPC message lets an attacker who has
compromised a lower-privilege process escape the sandbox and escalate
privilege. We restrict attention to the case where a lower-privilege
process directs IPC at a higher-privilege one, since the reverse
direction yields the attacker nothing, and distinguish three targets.

\noindent\textbf{Native JavaScript engine.} The JavaScript engine
runs inside a strict sandbox within the renderer. Memory corruption
within the engine affects only the engine's own sandboxed memory and
cannot, by itself, corrupt the renderer memory outside the sandbox. 
To escape, an attacker must first achieve code execution inside 
the engine sandbox,
then craft native JavaScript methods that write outside 
the sandbox region, thereby crossing the first privilege boundary. 
Only out of the sandbox write 
primitives are considered harmful in this threat model.

\noindent\textbf{Browser-process IPC.} Browser components communicate
over IPC. Here an attacker with full control of a lower-privilege
process crafts malicious IPC messages aimed at a higher-privilege one
(\eg renderer to GPU, or GPU to browser). If the receiver validates
its input insufficiently, the crafted message corrupts its memory,
converting code execution in a low-privilege process into memory
corruption in a high-privilege one.

\noindent\textbf{System calls.} Like any program, browser processes
issue system calls to the OS kernel. Lower-privilege processes are
restricted to a narrow set, higher-privilege processes to a broader
one. An attacker who has compromised the renderer can issue the
system calls available to it directly against the kernel, corrupting
kernel memory without traversing the intermediate browser-process
hierarchy at all~\cite{chrome_renderer_to_kernel}.

\begin{takeaway}
Attackers with a compromised low privilege process can 
issue crafted IPC calls to higher privilege components, 
corrupt their memory, and achieve privilege escalation.
\end{takeaway}

\definecolor{covR}{HTML}{E57373}
\definecolor{covO}{HTML}{FFB74D}
\definecolor{covY}{HTML}{FFE082}
\definecolor{covG}{HTML}{AED581}
\definecolor{covGG}{HTML}{66BB6A}
\definecolor{covNA}{HTML}{E0E0E0}
\begin{table*}[ht]
\centering
\caption{
Joint Privilege~$\times$~Input view of the \datasetMC
memory-corruption reports. 
The \emph{Gap?} column flags classes whose
bug-densest subtree sits below $30\%$ coverage.
}
\label{tab:cov-bug-gap}
\small
\setlength{\tabcolsep}{4pt}
\begin{tabular}{l rrrr r r l r l c}
\toprule
       & \multicolumn{4}{c}{\emph{Bug discovery by privilege}} & \emph{Total} & \emph{High-priv} & \multicolumn{2}{c}{\emph{Best-covered subtree}} & \emph{Bug-densest subtree} & \emph{Gap?} \\
\cmidrule(lr){2-5}\cmidrule(lr){9-10}
Class    & Low      & Mod.\    & High     & Kern.\   &        & share & name                          & cov.\,\% & name and cov.\,\%            &         \\
\midrule
Binary   & \hE{135} & \hA{9}   & \hB{25}  & \hA{1}   & 170    & $15\%$ & OSS-Fuzz parsers              & $60$--$90$ & device brokers $1$--$12$       & --      \\
Document & \hE{773} & \hE{126} & \hE{151} & \hA{1}   & 1{,}051 & $14\%$ & Blink (whole)                 & $36.6$    & \texttt{modules/} $0.1$--$5.5$ & \cellcolor{covR}\textbf{yes} \\
Script   & \hE{317} & \hC{54}  & \hZ{0}   & \hZ{0}   & 371    & $0\%$ & V8 (engine)                   & $64.7$    & shader back end (off-tree)     & \cellcolor{covR}\textbf{yes} \\
UI       & \hZ{1}   & \hZ{2}   & \hE{280} & \hZ{0}   & 283    & $99\%$ & whole-browser harness         & $14$--$34$ & no dedicated UI fuzzer         & \cellcolor{covR}\textbf{yes} \\
IPC      & \hD{73}  & \hD{68}  & \hE{174} & \hC{43}  & 358    & $61\%$ & \texttt{mojo}+\texttt{ipc}    & $55$      & \texttt{*/browser/} $15$--$28$ & \cellcolor{covR}\textbf{yes} \\
\bottomrule
\end{tabular}
\end{table*}

\section{Bug Discovery vs.\ Testing Coverage}
\label{sec:bugcov}
To answer RQ2, we identify the components where the discovered 
vulnerabilities arise and measure how well vendor testing covers them.
Specifically, we
map \datasetMC reported memory corruption vulnerabilities from
the past decade (2016--2025) onto the
Input~$\times$~Component~$\times$~Privilege taxonomy of
\autoref{sec:threat-model}, measure the line coverage Chromium
in-tree fuzzers achieve, and assess whether bug-manifesting
components are well tested.
\autoref{tab:cov-bug-gap} presents an overview.

\textbf{Data sources.} 
We collect vulnerability reports from the 
Chromium issue tracker and Firefox security advisories
~\cite{chrome_issue_tracker,firefox_security_bulletin}, 
from 2016-01-01 to 2025-12-01, including 
\chromeTotal Chrome reports and \firefoxreports Firefox 
reports\footnote{Firefox Bugzilla has a longer disclosure period,
so we fetched reports up to MFSA-2025-51.}.
We exclude Safari, whose tracker omits bug descriptions. 
Overall, our dataset of \totalreports bugs includes \datasetMC memory corruption reports. 
Moreover, we collect fuzzer code coverage from Chromium 
coverage dashboard~\cite{chromium_cov_dashboard}, and 
include the third-party library fuzzing coverage from the 
OSS-Fuzz coverage~\cite{ossfuzz_introspector} for third-party
parsers outside the Chromium tree. Line coverage presented  
are the sum of all fuzzing engines, including libFuzzer, Centipede 
and Fuzzilli. 

\textbf{Data processing.}
We collect five fields from the bug report: 
(i) whether the report describes a \emph{memory corruption} bug; (ii) the \emph{triggering input class};
(iii) the \emph{component name} 
where the bug manifests; (iv) the \emph{privilege level} of 
the process whose memory is corrupted;
and (v) the \emph{disclosure year}. 
Among all these fields, (iii) - (v) are processed using string 
matching and never go through the model. Specifically, we match 
the shepherd-classified module name to the component, then obtain the 
process privilege according to the component name.
The first two 
require reading the report context and developer response, 
so we extract them with \texttt{claude-opus-4-7}. 
To validate the labels, one author (who reported over 10
confirmed Chrome vulnerabilities) manually reviewed both
LLM-produced fields on a random sample of 100 reports, and
all 100 matched the LLM output.

\subsection{Binary Blob: Mostly Least Privilege, Parsers Well Covered}
\label{ssec:bugcov-binary}

Binary blobs are the only input class whose
testing investment is well aligned with the discovered bugs.
By design, every binary blob parser should run inside a
low-privilege process, following the principle of least
privilege~\cite{chrome_least_privileg,chrome_rule_of_2}, and
79\% (135/170) of the reports align with this principle.
The image, font, video, markup, and compression parsers
all are located inside a low privileged renderer process. 
Coverage mirrors the bug distribution: Most parsers, covered by 
OSS-Fuzz, achieve 50--90\% line-coverage, with only 
four below 50\% (\texttt{libwebp} 48, \texttt{libavif} 49,
\texttt{icu} 48, WebRTC 33).

The 21\% of bugs that impact privileged code concentrate
in a few components. The Firefox NSS cryptographic stack contributes
seven high-privilege reports that corrupt the parent process
directly, with no Chrome counterpart. Network-packet parsing
(Necko/HTTP and QUIC) contributes another five privileged 
reports. One ChromeOS Bluetooth-stack report reaches kernel
memory through the in-kernel BlueZ driver. The coverage of
privileged binary surfaces is uneven: SQLite (79\%),
libphonenumber (57\%), PAK (50\%), and NSS (43\%) are
reasonably fuzzed, while the device brokers remain
insufficiently tested (Bluetooth 1\%, WebHID 9\%, gdk-pixbuf
12\%).

\begin{takeaway}
Binary blobs are the only class where testing investment
matches the discovered bugs. Privileged components that violate least privilege are also poorly covered.
\end{takeaway}

\subsection{Documents: Bugs in WebAPI, Coverage in the Document Body}
\label{ssec:bugcov-doc}

\begin{table}[t]
\centering
\caption{Document class: coverage vs.\ bug count per surface.}
\label{tab:bugcov-doc}
\small
\setlength{\tabcolsep}{4pt}
\begin{tabular}{l r r}
\toprule
Surface                                         & Cov\,\% & Bugs \\
\midrule
\multicolumn{3}{l}{\emph{Blink (HTML)}} \\
\quad Document (\texttt{renderer/core/})        & 41.5 & 160 \\
\quad DocumentAPI (\texttt{renderer/core/})     & 41.5 & 240 \\
\quad WebAPI (\texttt{renderer/modules/})       & 17.4 & 485 \\
\midrule
\multicolumn{3}{l}{\emph{PDFium (PDF)}} \\
\quad \texttt{core}                             & 62.7 &  22 \\
\quad \texttt{xfa}                              & 56.8 &  74 \\
\quad \texttt{fpdfsdk}                          &  9.7 &  34 \\
\quad others (\texttt{third\_party/}, \dots)    &  --  &  36 \\
\bottomrule
\end{tabular}
\end{table}

Documents are the largest single input class in the dataset:
885 HTML reports (606 Chrome, 279 Firefox) and 166
Chrome PDF reports. The HTML reports partition cleanly into
three sub-categories.
The \emph{document body} (160 reports: parser, layout, CSS,
paint, SVG) is well sandboxed: 91\% stays in the renderer.
The \emph{document API} (240 reports: \texttt{window.*},
\texttt{document.*}, navigation, extensions) is also
renderer-dominated, with only 16\% escaping to UI/browser
processes.
The \emph{WebAPI bindings} (485 reports: WebGL, WebGPU,
WebRTC, Canvas, WebAudio, Media, Payments) is the sub-category
that drives bugs out of the renderer: 46\% of WebAPI reports
land in a moderate-, high-, or kernel-privilege process,
with WebGL alone contributing 107 reports against the GPU
process. PDF reports follow a different trajectory: all 166
remain at low privilege because PDFium's embedded JavaScript
cannot issue IPC.

The coverage profile is the inverse. Blink as a whole sits at
36.6\% line coverage. The document body parsers and the 
document API components (\texttt{renderer/core/}) reach 42--57\%.
The WebAPI surface (\texttt{renderer/modules/}) sits at
17\% aggregate, with the four bug-densest modules at
0.1\% (WebGL), 3.6\% (WebGPU) and 5.5\% (WebRTC). 
PDFium tells the same story at smaller
scale: the parsers are well covered, while the JavaScript-driven
SDK boundary \texttt{fpdfsdk} sits at 10\% (\autoref{tab:bugcov-doc}).

\begin{takeaway}
The document class shows the sharpest misalignment between
bugs and coverage. JavaScript-driven API bindings produce
most bugs yet receive the least coverage.
\end{takeaway}

\subsection{Scripts: Well Sandboxed, WebGL Backend Overlooked}
\label{ssec:bugcov-script}

\begin{table}[t]
\centering
\caption{Script class: coverage vs.\ bug count per surface.}
\label{tab:joint-script}
\small
\setlength{\tabcolsep}{4pt}
\begin{tabular}{l r r}
\toprule
Surface                                                      & Cov\,\% & Bugs \\
\midrule
JavaScript engine (\texttt{//v8/})                           & 64.7 & 312 \\
WebGL backend (\texttt{//third\_party/angle/})               & 34.0 &  11 \\
WebGPU backend (\texttt{//third\_party/dawn/})               & 73.6 &  13 \\
\bottomrule
\end{tabular}
\end{table}

The script class is the best-sandboxed class in the dataset:
no Script bug corrupts a browser-process or the kernel.
Among the 371 reports, 312 land at low privilege inside
the V8 sandbox region of the renderer (244 native JavaScript
and 68 WebAssembly). The remaining 59 shader reports split
into 54 in the moderate-privilege GPU process and 5 in the
renderer-side shader preprocessing.

Coverage is split cleanly by surface (\autoref{tab:joint-script}):
V8 and Dawn (WebGPU) are well covered, while ANGLE (WebGL) only
achieves 34\%.
Beyond the in-tree translators, the bulk of shader bugs
sits in stages that are not on the coverage dashboard at all:
DXC (the shader compiler that handles the ANGLE/Dawn output) 
accounts for
14 bugs in the dataset and lives outside the Chromium tree, 
other vendor compilers contribute 21 bugs
and live outside OSS-Fuzz. 
As these compilers are vendor specific and some are even close-sourced, 
we cannot measure their actual coverage~\cite{metal_close_source}.

\begin{takeaway}
The script class produces no high-privilege bugs, matching
the threat model. The JavaScript/Wasm engine and the WebGPU
translator are well-fuzzed, while the WebGL translator and
the vendor-controlled shader compilers are not.
\end{takeaway}

\subsection{UI Gestures: High-Privilege Concentrator, MiraclePtr Mitigated}
\label{ssec:bugcov-ui}

\begin{table}[t]
\centering
\caption{UI class: coverage vs.\ bug count per surface.}
\label{tab:joint-ui}
\small
\setlength{\tabcolsep}{4pt}
\resizebox{\columnwidth}{!}{%
\begin{tabular}{l r r}
\toprule
Surface (source directory)                            & Cov\,\% & Bugs \\
\midrule
BrowserChrome (\texttt{chrome/browser/ui})            & 14.3 & 201 \\
PlatformIntegration (\texttt{//ui/})                  & 33.8 &  72 \\
DevToolsWebUI (\texttt{chrome/browser/devtools})      &  2.4 &  10 \\
\bottomrule
\end{tabular}
}
\end{table}

UI gestures are the most privilege-concentrated class among five 
input classes: 99\% of the 283 UI reports corrupt the
high-privilege process directly, and ${\sim}$90\% of
Chrome UI reports are use-after-free. 
The bugs fall into three surfaces (\autoref{tab:joint-ui}):
\emph{BrowserChrome} (cross-platform widgets),
\emph{PlatformIntegration} (OS-specific toolkit), and
\emph{DevToolsWebUI} (internal pages).

The Chrome UI bug count climbs from 2 in 2016 to a peak of
112 in 2022, then collapses to 47, 22, 7, respectively. 
The reduction matches the deployment of 
MiraclePtr~\cite{chrome_miracle_ptr}, which keeps freed memory
alive while any \texttt{raw\_ptr<T>} reference exists, so any UAF
on a protected object becomes an unexploitable functional bug~\cite{chrome_severity_guideline}. 

Coverage of the UI class is the weakest of the five. The only
specialised fuzzers target accessibility features and lift
\texttt{ui/accessibility} and
\texttt{content/browser/accessibility}~\cite{chrome_ui_fuzz}. 
Everything else is covered as a side effect of whole-browser
harnesses (\autoref{tab:joint-ui}).

\begin{takeaway}
UI carries the highest percentage of high-privilege bugs
and the lowest dedicated coverage. A deployed mitigation
reduced the number of exploitable bugs, but the underlying
defects persist and the testing surface remains unaddressed.
\end{takeaway}

\subsection{IPC: Privilege Concentrator, Adapter Covered, Dispatch Not}
\label{ssec:bugcov-ipc}

\begin{table}[t]
\centering
\caption{IPC class: coverage vs.\ bug count per surface.}
\label{tab:joint-ipc}
\small
\setlength{\tabcolsep}{4pt}
\begin{tabular}{l r r}
\toprule
Surface (directory)                  & Cov\,\% & Bugs \\
\midrule
\multicolumn{3}{l}{\emph{Receiver}} \\
\texttt{chrome/browser/}             & 15.3 & 45 \\
\texttt{content/browser/}            & 27.7 & 44 \\
\texttt{components/}                 & 20.3 & 13 \\
\texttt{services/}                   & 36.2 & 10 \\
\midrule
\multicolumn{3}{l}{\emph{Adapter}} \\
\texttt{mojo/}                       & 55.4 & \multirow{2}{*}{7} \\
\texttt{ipc/}                        & 55.1 &     \\
\bottomrule
\end{tabular}
\end{table}

IPC is the only class that spans all four privilege
tiers, and the only class that targets the privilege boundaries. 
The 358 reports partition
into three sub-types. \emph{V8 sandbox escape} (73 Chrome
reports) corrupts memory inside the V8 sandbox region. It is
low-privilege at the point of corruption but routes to a
sandbox-escape outcome. \emph{Normal IPC} 
(182 Chrome and 25 Firefox reports) is the canonical
sandbox-escape vector: 173 (84\%) land in a high-privilege
host process and 34 in moderate-privilege utility, GPU, or
network processes. \emph{Syscall IPC} (78 reports) reaches the
kernel directly: 43 corrupt kernel memory, 34 land in the
\emph{virglrenderer} (a GPU process handling VM requests on
ChromeOS), and one reaches the ChromeOS parent process, which
has higher privilege than the normal browser process.

Coverage flips between the IPC adapter (where IPC requests 
are forwarded) and the dispatch targets (where IPC requests are 
processed). 
The adapter (\texttt{mojo/}, \texttt{ipc/}) sits at
55\% but accounts for only 7 historical reports.
In contrast, the dispatch targets manifest the remaining 112
vulnerabilities at much lower coverage (\autoref{tab:joint-ipc}):
\texttt{chrome/browser/} and \texttt{content/browser/} together
own 75\% of the in-tree IPC bugs and run in the high-privilege
browser process. \texttt{services/<svc>/} is the only receiver
tier whose modal privilege is moderate, because each service
launches into its own utility process.

\begin{takeaway}
IPC's coverage is the inverse of where its bugs are. The
well-covered adapter holds few bugs, while the receiver 
that hold most bugs are poorly covered.
\end{takeaway}

\subsection{Vulnerability Trend In The Past Decade}
\label{ssec:bugcov-timeline}

\begin{figure}[t]
\centering
\begin{tikzpicture}
\begin{axis}[
    name=cbtplot,
    width=\columnwidth, height=3.6cm,
    enlarge x limits={abs=0.45},
    ymin=0,
    xtick={2016,2017,2018,2019,2020,2021,2022,2023,2024,2025},
    xticklabels=\empty,
    scaled x ticks=false,
    yticklabel style={font=\footnotesize},
    ylabel={\#N reports},
    ylabel style={font=\footnotesize, yshift=-6pt},
    grid=major, grid style={draw=black!12},
    axis line style={draw=black!55},
    every axis plot/.append style={line width=0.7pt, mark size=1.3pt},
    clip=false,
]
\addplot[color=clsBinary, mark=*]    coordinates {(2016,68)(2017,30)(2018,11)(2019,11)(2020,9)(2021,6)(2022,7)(2023,15)(2024,11)(2025,2)};
\addplot[color=clsDoc, mark=square*] coordinates {(2016,110)(2017,109)(2018,101)(2019,92)(2020,95)(2021,125)(2022,145)(2023,135)(2024,86)(2025,53)};
\addplot[color=clsScript, mark=triangle*] coordinates {(2016,19)(2017,22)(2018,22)(2019,32)(2020,20)(2021,25)(2022,26)(2023,42)(2024,109)(2025,54)};
\addplot[color=clsUI, mark=diamond*] coordinates {(2016,3)(2017,3)(2018,3)(2019,8)(2020,10)(2021,59)(2022,115)(2023,50)(2024,25)(2025,7)};
\addplot[color=clsIPC, mark=otimes*] coordinates {(2016,3)(2017,9)(2018,5)(2019,18)(2020,31)(2021,59)(2022,41)(2023,93)(2024,37)(2025,62)};
\end{axis}

\tikzset{yrlbl/.style={font=\scriptsize, anchor=east, inner sep=1pt}}
\foreach \frac/\yr/\mc/\allv in {%
    0.0455/2016/203/376,
    0.1465/2017/173/351,
    0.2475/2018/142/354,
    0.3485/2019/161/347,
    0.4495/2020/165/349,
    0.5505/2021/274/452,
    0.6515/2022/334/553,
    0.7525/2023/335/588,
    0.8535/2024/268/444,
    0.9545/2025/178/285}
{
    \node[yrlbl] at ($(cbtplot.south west)!\frac!(cbtplot.south east) + (7pt,-9pt)$)  {\yr};
    \node[yrlbl] at ($(cbtplot.south west)!\frac!(cbtplot.south east) + (7pt,-19pt)$) {\mc};
    \node[yrlbl] at ($(cbtplot.south west)!\frac!(cbtplot.south east) + (7pt,-29pt)$) {\allv};
}
\node[font=\scriptsize, anchor=east] at ($(cbtplot.south west)+(-2pt,-9pt)$)  {Year};
\node[font=\scriptsize, anchor=east] at ($(cbtplot.south west)+(-2pt,-19pt)$) {MC};
\node[font=\scriptsize, anchor=east] at ($(cbtplot.south west)+(-2pt,-29pt)$) {All};
\draw[black!55, line width=0.4pt]
    ($(cbtplot.south west)+(0,-5pt)$) -- ($(cbtplot.south east)+(0,-5pt)$);
\draw[black!25, line width=0.3pt]
    ($(cbtplot.south west)+(0,-13pt)$) -- ($(cbtplot.south east)+(0,-13pt)$);
\draw[black!55, line width=0.4pt]
    ($(cbtplot.south west)+(0,-33pt)$) -- ($(cbtplot.south east)+(0,-33pt)$);

\node[anchor=north, font=\scriptsize, inner sep=1pt]
     at ($(cbtplot.south)+(0,-37pt)$) {%
    \textcolor{clsBinary}{$\bullet$}~Binary\quad
    \textcolor{clsDoc}{$\blacksquare$}~Document\quad
    \textcolor{clsScript}{$\blacktriangle$}~Script\quad
    \textcolor{clsUI}{$\blacklozenge$}~UI\quad
    \textcolor{clsIPC}{$\bullet$}~IPC%
};
\end{tikzpicture}
\caption{Memory corruption and total vulnerability reports by year and input class.
MC: Memory corruption.}
\label{fig:cov-bug-time}
\end{figure}
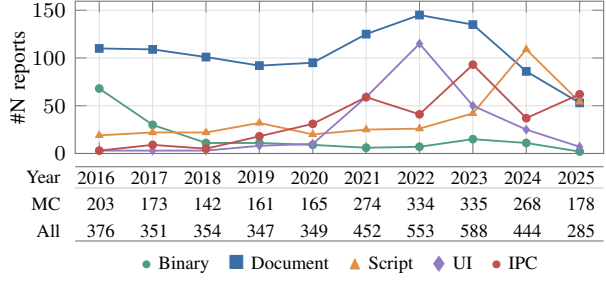

The per-year decomposition in \autoref{fig:cov-bug-time} shows
that each class follows a distinct trajectory, from which we
conclude four shifts.

\textit{Deployed mitigations reshape the bug-discovery
distribution.}
UI disclosures climbed to a peak in 2022 and collapsed after
the MiraclePtr deployment (\autoref{ssec:bugcov-ui}),
demonstrating that a single deployed mitigation can collapse a
bug class without big code changes.

\textit{Upstream features introduce new bugs.}
WebAssembly (a subclass of script) discoveries remained in 
the single digits through
2023 and stepped up to 37 in 2024. The step coincides
with the shipping of WebAssembly garbage collection, the
JavaScript promise integration, exception handling, and
relaxed SIMD~\cite{chrome_ship_wasm_gc,chrome_ship_wasm_jspi,chrome_ship_wasm_simd},
a set of new features landing in 2023 and 2024. 
The implication is direct: under-tested new code introduces new vulnerabilities.

\textit{A new harness on a previously-untested surface
boosts the bug discovery.}
Shader bugs spiked in 2024: of the 59 shader reports across
the decade, 33 land in that year alone, mostly from one
academic work~\cite{bernhard2024darthshader} that built new
harnesses for the WebGPU translator stack. A new harness on an
untested surface boosts a class's bug discovery at a much
smaller effort than improving fuzzing across the whole browser.

\textit{Discovered bugs are shifting from in-renderer to
cross-privilege surfaces.}
IPC bugs increase from 3 in 2016 to a peak of 93 in
2023, while binary-blob reports collapse from 68 in 2016 to
single digits after 2018 as the Flash adapter retires and
OSS-Fuzz improves. The shallow in-renderer surfaces are now
well tested, and attack targets are moving from low-privilege
to high-privilege code.

\begin{takeaway}
  The binary-blob and UI surfaces recede, while the IPC, Wasm,
  and shader surfaces grow.
\end{takeaway}

\section{Browser Testing Techniques}
\label{sec:browser-testing}

To answer RQ3, we examine whether browser testing techniques 
cover the bug-dense attack surfaces. We primarily focus on
the browser fuzzing techniques and discuss other testing 
techniques in~\autoref{sec:discuss}.
This section first decomposes the fuzzing pipeline,
then surveys academic browser fuzzers along the attack surfaces,
and finally identifies overlooked deployment gaps.

\newcommand{\ax}{\ding{108}}
\newcommand{\axN}{\textcolor{black!15}{\ding{109}}}

\definecolor{covBlue}{HTML}{1F4E8C}
\definecolor{bugOrange}{HTML}{C2691F}
\definecolor{vulnRed}{HTML}{A21F1F}
\definecolor{bntGreen}{HTML}{1F7A36}
\newcommand{\axC}{\textcolor{covBlue}{\faChartLine}}
\newcommand{\axCN}{\textcolor{black!15}{\faChartLine}}
\newcommand{\axB}{\textcolor{bugOrange}{\faBug}}
\newcommand{\axBN}{\textcolor{black!15}{\faBug}}
\newcommand{\axV}{\textcolor{vulnRed}{\faBomb}}
\newcommand{\axVN}{\textcolor{black!15}{\faBomb}}
\newcommand{\axD}{\textcolor{bntGreen}{\textbf{\$}}}
\newcommand{\axDN}{\textcolor{black!15}{\textbf{\$}}}

\begin{table*}[t]
\centering
\caption{Browser-relevant fuzzing techniques targeting
\emph{binary} and \emph{document} inputs.
\textbf{M} mutation technique, \textbf{G} generation technique,
\textbf{F} intermediate feedback, \textbf{O} bug oracle, and
\textbf{D} deployment.
Four further columns stand for the paper's effectiveness claims:
\textbf{Cov} -- paper claims a coverage improvement;
\textbf{Bug} -- paper claims to find new bugs in the target;
\textbf{Brow} -- paper claims to find new browser vulnerabilities;
\textbf{\$} -- paper receives bounty from browser vendors;
\textbf{Bug} is different from \textbf{Brow}, as fuzzers may target 
non-browser implementation (\eg COOPER~\cite{xu2022cooper} tests 
Adobe and Foxit PDF).
}
\label{tab:fuzzer-axes-a}
\footnotesize
\setlength{\tabcolsep}{3pt}
\resizebox{\linewidth}{!}{
\begin{tabular}{l l l c c c c c c c c c l}
\toprule
                      &      &         & \multicolumn{5}{c}{Contributions} & \multicolumn{4}{c}{Effectiveness} &                       \\
\cmidrule(lr){4-8}\cmidrule(lr){9-12}
Technique             & Year & Venue   & M    & G    & F    & O    & D    & Cov  & Bug  & Brow & \$   & One-line summary of core contribution \\
\midrule
\multicolumn{13}{l}{\textit{Binary-blob fuzzers — general-purpose CGFs; target third-party parsers in the browser}} \\
AFLFast~\cite{bohme2016coverage}          & 2016 & CCS       & \ax  & \axN & \axN & \axN & \axN & \axC  & \axB  & \axVN  & \axDN & Markov-chain seed scheduling \\
Driller~\cite{stephens2016driller}        & 2016 & NDSS      & \ax  & \axN & \axN & \axN & \axN & \axC  & \axB  & \axVN  & \axDN & Concolic fallback when CGF stalls \\
AFLGo~\cite{bohme2017directed}            & 2017 & CCS       & \ax  & \axN & \ax  & \axN & \axN & \axCN & \axB  & \axVN  & \axDN & Distance-to-target directed fuzzing \\
Angora~\cite{chen2018angora}              & 2018 & S\&P      & \ax  & \axN & \axN & \axN & \axN & \axC  & \axB  & \axVN  & \axDN & Byte-level taint + gradient descent \\
CollAFL~\cite{gan2018collafl}             & 2018 & S\&P      & \axN & \axN & \ax  & \axN & \axN & \axC  & \axB  & \axVN  & \axDN & Path-sensitive coverage resolves hash collisions \\
QSYM~\cite{yun2018qsym}                   & 2018 & USENIX    & \ax  & \axN & \axN & \axN & \axN & \axC  & \axB  & \axVN  & \axDN & Fast hybrid symbolic execution \\
Redqueen~\cite{aschermann2019redqueen}    & 2019 & NDSS      & \ax  & \axN & \ax  & \axN & \axN & \axC  & \axB  & \axVN  & \axDN & Input-to-state correspondence; register values as feedback \\
EcoFuzz~\cite{yue2020ecofuzz}             & 2020 & USENIX    & \ax  & \axN & \axN & \axN & \axN & \axC  & \axB  & \axVN  & \axDN & Energy allocation under adversarial bandits \\
GreyOne~\cite{gan2020greyone}             & 2020 & USENIX    & \ax  & \axN & \axN & \axN & \axN & \axC  & \axB  & \axVN  & \axDN & Taint-aided byte mutation \\
ParmeSan~\cite{osterlund2020parmesan}      & 2020 & USENIX    & \ax  & \axN & \ax  & \axN & \axN & \axC  & \axB  & \axVN  & \axDN & Sanitizer-guided distance-to-target feedback \\
SymCC~\cite{poeplau2020symbolic}          & 2020 & USENIX    & \ax  & \axN & \axN & \axN & \axN & \axC  & \axB  & \axVN  & \axDN & Compile-time symbolic-execution instrumentation \\
TortoiseFuzz~\cite{wang2020not}           & 2020 & NDSS      & \axN & \axN & \ax  & \axN & \axN & \axCN & \axB  & \axVN  & \axDN & Security-instruction-weighted feedback \\
Weizz~\cite{fioraldi2020weizz}            & 2020 & ISSTA     & \ax  & \axN & \axN & \axN & \axN & \axC  & \axB  & \axVN  & \axDN & Automatic format inference \\
OptiMin~\cite{herrera2021seed}            & 2021 & ISSTA     & \ax  & \axN & \axN & \axN & \axN & \axCN & \axBN & \axVN  & \axDN & Empirical study of seed-selection heuristics \\
Jigsaw~\cite{chen2022jigsaw}              & 2022 & S\&P      & \ax  & \axN & \axN & \axN & \axN & \axC  & \axB  & \axVN  & \axDN & Efficient path-constraint solver for CGF \\
K-Scheduler~\cite{she2022effective}       & 2022 & S\&P      & \ax  & \axN & \ax  & \axN & \axN & \axC  & \axB  & \axVN  & \axDN & Reachable-node count as feedback for seed scheduling \\
PATA~\cite{liang2022pata}                 & 2022 & S\&P      & \ax  & \axN & \axN & \axN & \axN & \axC  & \axB  & \axVN  & \axDN & Path-aware taint analysis \\
SymSan~\cite{chen2022symsan}              & 2022 & USENIX    & \ax  & \axN & \axN & \axN & \axN & \axC  & \axB  & \axVN  & \axDN & Sanitizer-style symbolic-execution backend \\
Truzz~\cite{zhang2022path}                & 2022 & ICSE      & \ax  & \axN & \axN & \axN & \axN & \axC  & \axB  & \axVN  & \axDN & Runtime program-state guided scheduling \\
AIFORE~\cite{shi2023aifore}               & 2023 & USENIX    & \ax  & \axN & \axN & \axN & \axN & \axC  & \axB  & \axVN  & \axDN & AI-driven format inference \\
FishFuzz~\cite{zheng2023fishfuzz}         & 2023 & USENIX    & \ax  & \axN & \ax  & \axN & \axN & \axC  & \axB  & \axVN  & \axDN & Many-target directed fuzzing with distance-based feedback \\
MendelFuzz~\cite{zheng2025mendelfuzz}     & 2025 & FSE       & \ax  & \axN & \axN & \axN & \axN & \axC  & \axB  & \axVN  & \axDN & Selective deterministic stage on critical bytes \\
ZTaint-Havoc~\cite{xie2025ztaint}         & 2025 & ISSTA     & \ax  & \axN & \axN & \axN & \axN & \axC  & \axB  & \axVN  & \axDN & Zero-execution taint inference in havoc \\
\addlinespace
\multicolumn{13}{l}{\textit{Document fuzzers (HTML / PDF); including the WebAPI and document APIs}} \\
FreeDom~\cite{xu2020freedom}              & 2020 & CCS       & \ax  & \ax  & \axN & \axN & \axN & \axC & \axB  & \axV  & \axD  & Context-aware IR; supports both generative and mutation modes \\
Favocado~\cite{dinh2021favocado}          & 2021 & NDSS      & \ax  & \ax  & \axN & \axN & \axN & \axCN & \axB  & \axV  & \axD  & IDL-driven generation + state-aware mutation \\
FuzzOrigin~\cite{kim2022fuzzorigin}       & 2022 & USENIX    & \axN & \axN & \axN & \ax  & \axN & \axCN & \axB  & \axV  & \axDN  & Static origin-tagging oracle \\
Minerva~\cite{zhou2022minerva}            & 2022 & FSE       & \axN & \ax  & \axN & \axN & \axN & \axC  & \axB  & \axV  & \axDN  & Mod-ref API graph guides API-sequence synthesis \\
GLeeFuzz~\cite{peng2023gleefuzz}          & 2023 & USENIX    & \ax  & \axN & \ax  & \axN & \axN & \axC & \axB  & \axV  & \axDN  & GL error messages as lightweight feedback for WebGL \\
COOPER~\cite{xu2022cooper}                & 2022 & NDSS      & \ax  & \axN & \axN & \axN & \axN & \axC  & \axB  & \axVN  & \axDN & Cooperative mutation of PDF + embedded JS \\
TypeOracle~\cite{guo2023operand}          & 2023 & ICSE      & \axN & \ax  & \axN & \axN & \axN & \axC & \axB  & \axVN  & \axDN & Differential operand-variation type inference \\
\bottomrule
\end{tabular}
}
\end{table*}

\begin{table*}[t]
\centering
\caption{Browser-relevant fuzzing techniques targeting
\emph{script}, \emph{shader}, and \emph{sandbox/IPC} inputs.
Legend matches \autoref{tab:fuzzer-axes-a}.}
\label{tab:fuzzer-axes-b}
\footnotesize
\setlength{\tabcolsep}{3pt}
\resizebox{\linewidth}{!}{
\begin{tabular}{l l l c c c c c c c c c l}
\toprule
                      &      &         & \multicolumn{5}{c}{Contributions} & \multicolumn{4}{c}{Effectiveness} &                       \\
\cmidrule(lr){4-8}\cmidrule(lr){9-12}
Technique             & Year & Venue   & M    & G    & F    & O    & D    & Cov  & Bug  & Brow & \$   & One-line summary of core contribution \\
\midrule
\multicolumn{13}{l}{\textit{JavaScript-engine fuzzers}} \\
CodeAlchemist~\cite{han2019codealchemist} & 2019 & NDSS      & \axN & \ax  & \axN & \axN & \axN & \axC  & \axB  & \axV  & \axDN  & AST code-bricks with assembly constraints \\
DIE~\cite{park2020fuzzing}                & 2020 & S\&P      & \ax  & \axN & \axN & \axN & \axN & \axC  & \axB  & \axV  & \axD  & Aspect-preserving mutation of typed AST \\
Montage~\cite{lee2020montage}             & 2020 & USENIX    & \axN & \ax  & \axN & \axN & \axN & \axCN  & \axB  & \axV  & \axD  & LSTM-trained AST fragment generation \\
Jest~\cite{park2021jest}                  & 2021 & ICSE      & \axN & \axN & \axN & \ax  & \axN & \axCN & \axB  & \axV  & \axDN  & N+1-version differential testing of JS engines \\
SOFI~\cite{he2021sofi}                    & 2021 & CCS       & \ax  & \ax  & \axN & \axN & \axN & \axC  & \axB  & \axV  & \axDN  & Reflection-augmented mutation and generation \\
JIT-Picking~\cite{bernhard2022jit}        & 2022 & CCS       & \axN & \axN & \axN & \ax  & \axN & \axCN & \axB  & \axV  & \axD  & Differential test between JIT tiers \\
Fuzzilli~\cite{gross2023fuzzilli}         & 2023 & NDSS      & \ax  & \ax  & \axN & \axN & \axN & \axC  & \axB  & \axV  & \axDN  & FuzzIL IR with mutation engine and code generators \\
FuzzJIT~\cite{wang2023fuzzjit}            & 2023 & USENIX    & \ax  & \ax  & \axN & \ax  & \axN & \axC  & \axB  & \axV  & \axDN  & JIT-tier-aware mutation atop Fuzzilli's generation \\
FuzzFlow~\cite{xu2024fuzzing}             & 2024 & CCS       & \ax  & \axN & \axN & \axN & \axN & \axC  & \axB  & \axV  & \axDN  & FlowIR graph-based IR mutation with decoupled control/data flow \\
OptFuzz~\cite{wang2024optfuzz}            & 2024 & USENIX    & \axN & \axN & \ax  & \axN & \axN & \axC  & \axB  & \axV  & \axDN  & JIT-optimisation-path feedback \\
Dumpling~\cite{wachter2025dumpling}       & 2025 & NDSS      & \axN & \axN & \axN & \ax  & \axN & \axCN & \axB  & \axV  & \axD & Fine-grained differential JIT oracle \\
\addlinespace
\multicolumn{13}{l}{\textit{WebAssembly fuzzers}} \\
WADIFF~\cite{zhou2023wadiff}              & 2023 & ASE       & \axN & \ax  & \axN & \ax  & \axN & \axCN  & \axB  & \axVN  & \axDN & Symbolic-execution-driven per-instruction generation from spec \\
Wapplique~\cite{zhao2024wapplique}        & 2024 & ISSTA     & \ax  & \axN & \axN & \axN & \axN & \axC  & \axB  & \axVN  & \axDN & Code-fragment appliqu\'e substitution into seed bytecode \\
Wasmaker~\cite{cao2024wasmaker}           & 2024 & ISSTA     & \ax  & \axN & \axN & \ax  & \axN & \axCN  & \axB  & \axVN  & \axDN & Semantic-aware disassemble-reassemble of real-world binaries \\
FreeWavm~\cite{qian2025freewavm}          & 2025 & ISSTA     & \ax  & \axN & \axN & \axN & \axN & \axC  & \axB  & \axVN  & \axDN & Parse-tree structure mutation with snapshot guidance \\
LWDIFF~\cite{zhou2025lwdiff}              & 2025 & ICSE      & \axN & \ax  & \axN & \ax  & \axN & \axC & \axB  & \axVN  & \axDN & LLM-extracted specification knowledge drives generation \\
RGFuzz~\cite{park2025rgfuzz}              & 2025 & S\&P      & \axN & \ax  & \axN & \ax  & \axN & \axC  & \axB  & \axV  & \axDN & Compiler-rule-guided generation with reverse-stack synthesis \\
Stanley'25~\cite{stanley2025finding}      & 2025 & ASE       & \axN & \ax  & \axN & \ax  & \axN & \axCN & \axB  & \axVN  & \axDN & Random differential testing of WIT binding generators \\
Waltzz~\cite{zhang2025waltzz}             & 2025 & USENIX    & \ax  & \ax  & \axN & \axN & \axN & \axC  & \axB  & \axVN  & \axDN & Stack-invariant mutators with skeleton-based snippet generation \\
WEST~\cite{youn2025west}                  & 2025 & ASE       & \axN & \ax  & \axN & \axN & \axN & \axCN & \axB  & \axV  & \axDN & SpecTec mechanized-specification-driven test generation \\
\addlinespace
\multicolumn{13}{l}{\textit{Shader fuzzers}} \\
DarthShader~\cite{bernhard2024darthshader}& 2024 & CCS       & \ax  & \axN & \axN & \axN & \ax  & \axC  & \axB  & \axV  & \axD  & Dual AST/IR mutation; WebGPU translator harness \\
\addlinespace
\multicolumn{13}{l}{\textit{Sandbox / IPC fuzzers}} \\
SbxBrk~\cite{bars2025empirical}           & 2025 & CCS       & \axN & \axN & \ax  & \axN & \axN & \axCN & \axB  & \axV  & \axD  & Sandbox-boundary read events as feedback \\
\bottomrule
\end{tabular}
}
\end{table*}

\subsection{Fuzzing Pipeline and Deployment}
\label{ssec:fuzzer-pipeline}

\begin{figure}[t!]
	\centering
  \includegraphics[width=\columnwidth]{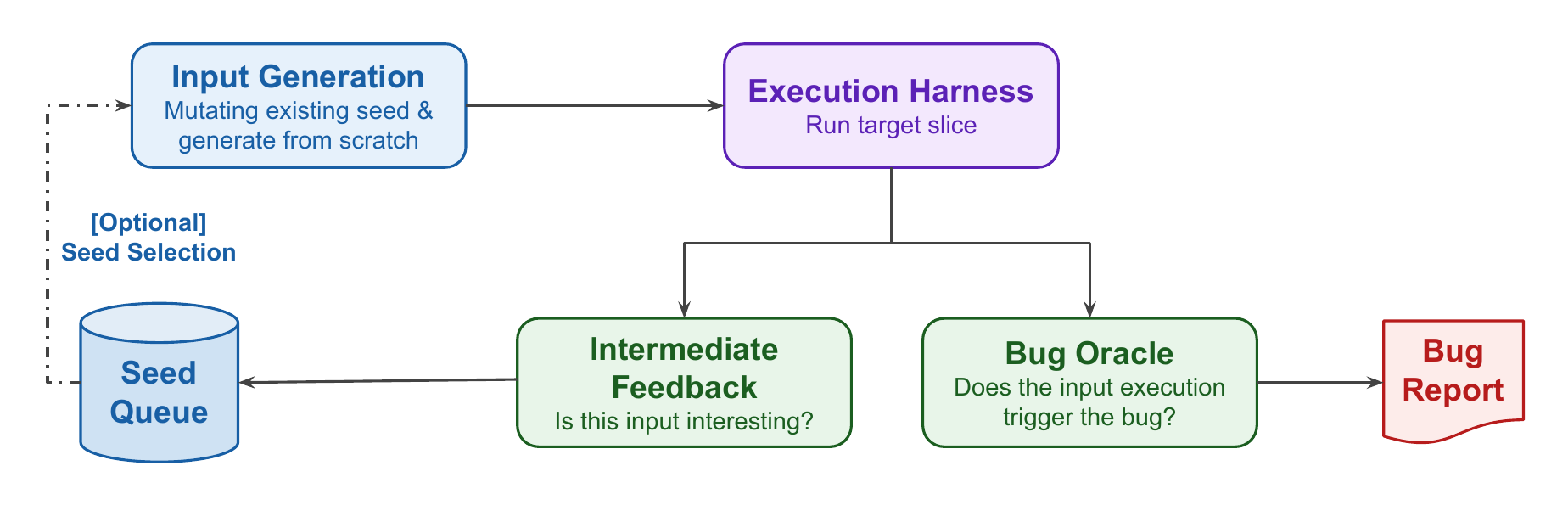}
  \caption{The pipeline of a fuzzer. }
  \label{fig:fuzz-pipeline}
\end{figure}

A fuzzer is built from five stages that together determine
which inputs it produces, how it judges progress, and how it is
wired into the target. These five stages map to the columns
of \autoref{tab:fuzzer-axes-a} and \autoref{tab:fuzzer-axes-b}:
\textbf{M}~mutation, \textbf{G}~generation, \textbf{F}~intermediate
feedback, \textbf{O}~bug oracle, and \textbf{D}~deployment.

\textbf{(M) Mutation technique.}
Mutation-based fuzzers derive each new input from an existing
seed through two internal stages: \emph{seed selection}
(which seed to mutate) and \emph{mutation} (how to mutate the seed). 
For structured inputs, mutation operates 
on a high-level representation (\eg AST or IR) so that syntactic
validity is preserved across edits.

\textbf{(G) Generation technique.}
Generation-based fuzzers synthesise each input from scratch
rather than mutating a seed. Specifically, the fuzzer
leverages a template (\eg grammar,
IDL, or custom IR) to generate inputs.
The selected template varies for distinct targets: 
context-free grammars for HTML/CSS, IDL files for browser API
bindings, and specifications for WebAssembly modules.

\textbf{(F) Intermediate feedback.}
Intermediate feedback is the signal that decides which inputs
are interesting enough to keep and re-mutate. The canonical
signal is edge coverage produced by instrumentation, 
but a long line of work refines or replaces it with a more
domain-specific signal: path-sensitive coverage,
memory operation coverage,
sandbox-boundary read events, GL error messages, register
values, call-graph distance, or reachable-node counts.

\textbf{(O) Bug oracle.}
The bug oracle is the predicate that decides whether a
particular execution exposed a bug. Most fuzzers rely on memory
sanitizers~\cite{serebryany2012addresssanitizer}, which cannot
detect logic bugs. Thus, many works implement differential
oracles to find vulnerabilities beyond memory corruption, 
by comparing the different execution results. 

\textbf{(D) Deployment.}
Deployment is how the fuzzer is connected to the target in
practice. It includes: (i)~\emph{harness design},
which decides the program slice the fuzzer actually drives
(whole browser, hand-written deep-API driver); and (ii)~\emph{fuzzer
configuration}, which decides how the available grammars,
harnesses, and feedback signals are composed in the running
fuzzer. Most academic works inherit the vendor's default
deployment (\autoref{ssec:fuzzer-missing}).

\subsection{Existing Fuzzers By Attack Surface}
\label{ssec:fuzzer-axes}

\autoref{tab:fuzzer-axes-a} and~\autoref{tab:fuzzer-axes-b} 
provide a survey of existing browser fuzzers,
grouped by their target attack surfaces.
For each surface, we first recall whether there is a gap 
between bug discovery and testing coverage (\autoref{sec:bugcov}), then
discuss whether the surveyed techniques can fill the gap.

\textbf{Binary-blob fuzzers.}
Most binary-blob bugs arise in low-privilege parsers, which are
already well tested (\autoref{ssec:bugcov-binary}).
As standalone libraries, these parsers are tested by
general-purpose fuzzers, which primarily leverage
mutation-based input generation. For
\emph{seed selection}, existing works model the fuzzing
process~\cite{bohme2016coverage,yue2020ecofuzz}, calculate
the seed-target distance~\cite{bohme2017directed,
she2022effective,zheng2023fishfuzz} and measure the impact of initial corpus~\cite{herrera2021seed}. For
\emph{mutation}, they introduce concolic
execution~\cite{stephens2016driller,yun2018qsym,
poeplau2020symbolic,chen2022jigsaw,chen2022symsan}, taint
analysis~\cite{chen2018angora,aschermann2019redqueen,
gan2020greyone,liang2022pata,xie2025ztaint}, and input-format
inference~\cite{fioraldi2020weizz,zhang2022path,shi2023aifore,
zheng2025mendelfuzz} to improve constraint solving. For
\emph{intermediate feedback}, a line of work replaces the
default edge-coverage signal with a domain-specific signal:
CollAFL~\cite{gan2018collafl} introduces path-sensitive
coverage that resolves hash collisions;
TortoiseFuzz~\cite{wang2020not} directs fuzzing toward code
that exercises memory-unsafe operations through customized
feedback;
Redqueen~\cite{aschermann2019redqueen} uses register values
observed at compare instructions as an additional input-to-state
feedback signal;
AFLGo~\cite{bohme2017directed},
ParmeSan~\cite{osterlund2020parmesan}, and
FishFuzz~\cite{zheng2023fishfuzz} use the call-graph distance
from each input to the target sites as feedback to steer
mutation; and K-Scheduler~\cite{she2022effective} uses the
number of reachable nodes as feedback to score seeds.
The bug oracle relies on memory
sanitizers~\cite{serebryany2012addresssanitizer}, and the
harness is typically the OSS-Fuzz library wrapper for each
codec or parser library.
The remaining gap, the poorly covered privileged parsers,
stems not from the techniques but from the deployment
(\autoref{ssec:fuzzer-missing}).

\textbf{Document fuzzers.}
The document class shows the sharpest gap: the WebAPI bindings
and the PDFium SDK boundary hold most bugs yet the least
coverage (\autoref{ssec:bugcov-doc}).
For document testing, byte-level mutation breaks
syntax validity, so works operate on high-level representations.
For \emph{mutation}, existing works propose cooperative mutation
of document objects together with their embedded
JavaScript~\cite{xu2022cooper} and feedback-guided structural
mutation of WebGL API call sequences driven by
domain-specific error signals~\cite{peng2023gleefuzz}.
For \emph{generation}, context-free grammars~\cite{domato} and
context-aware IRs~\cite{xu2020freedom} synthesise HTML
documents, mod-ref API dependency graphs compose call sequences
with high inter-API interaction~\cite{zhou2022minerva},
vendor-defined IDLs collect candidate
APIs~\cite{dinh2021favocado}, and differential type inference
recovers parameter types when the IDL is
incomplete~\cite{guo2023operand}.
For the \emph{bug oracle}, beyond memory sanitizers, a static
origin-tagging oracle has been proposed to detect
non-memory-corruption UXSS bugs~\cite{kim2022fuzzorigin}. The
\emph{intermediate feedback} signal is either the edge
coverage or the GL error messages~\cite{peng2023gleefuzz}.
Several document fuzzers already target the bug-dense
WebAPI bindings~\cite{zhou2022minerva, dinh2021favocado,
guo2023operand,domato}. However, vendor deployments
fail to import the available grammars, which leaves these
modules under-covered. The PDFium boundary suffers instead
from an oversimplified harness (\autoref{ssec:fuzzer-missing}).

\textbf{Script fuzzers (JavaScript, WebAssembly, shaders).}
In the script class, the JavaScript/Wasm engines are well
fuzzed. The gap sits at the WebGL backend and the vendor
shader compilers (\autoref{ssec:bugcov-script}).
As JavaScript, Wasm, and shaders are all strictly validated 
before execution, works raise the mutation granularity above 
bytes to preserve validity.
For \emph{mutation}, existing JavaScript and shader works
operate on typed ASTs to preserve language
aspects~\cite{park2020fuzzing,bernhard2024darthshader} or on
graph-based IRs that decouple control and data
flow~\cite{xu2024fuzzing}; existing Wasm works substitute
code fragments into seed bytecode~\cite{zhao2024wapplique},
mutate parse-tree structure with snapshot
guidance~\cite{qian2025freewavm}, and synthesise
stack-invariant instruction snippets~\cite{zhang2025waltzz}.
For \emph{generation}, JavaScript fuzzers compose programs
from semantics-aware code bricks~\cite{han2019codealchemist},
learned language models~\cite{lee2020montage}, and
reflection-discovered runtime types~\cite{he2021sofi}, and
further specialise an IR to the JIT
compiler~\cite{gross2023fuzzilli,wang2023fuzzjit}. Wasm
fuzzers synthesise stack-valid modules driven by spec-to-DSL
symbolic execution~\cite{zhou2023wadiff}, mechanised
specifications~\cite{youn2025west}, LLM-extracted spec
knowledge~\cite{zhou2025lwdiff}, production-runtime rewrite
rules~\cite{park2025rgfuzz}, and skeleton-based snippet
templates~\cite{zhang2025waltzz}.
For \emph{intermediate feedback}, JS engine fuzzers refine
the default edge-coverage signal by tracking JIT optimisation
paths~\cite{wang2024optfuzz} and sandbox-boundary memory
reads~\cite{bars2025empirical}.
For the \emph{bug oracle}, differential testing surfaces
non-crashing semantic bugs: JIT-vs-interpreter comparison
within a single engine~\cite{bernhard2022jit,
wachter2025dumpling,wang2023fuzzjit}, specification-as-oracle
across multiple engine implementations~\cite{park2021jest},
cross-runtime comparison on real-world Wasm
binaries~\cite{cao2024wasmaker}, and
cross-binding-generator divergence on WIT
interfaces~\cite{stanley2025finding}.
Most works thus further harden the well-fuzzed engines. Only
DarthShader~\cite{bernhard2024darthshader} builds a dedicated
harness for the WebGPU shader-translation pipeline (Dawn),
which sits deep in the browser and is hard to reach. The WebGL
backend (ANGLE) and the vendor shader compilers remain
under-tested.

\textbf{UI-gesture and IPC fuzzers.}
Bugs in UI gestures (\autoref{ssec:bugcov-ui}) and IPC calls
(\autoref{ssec:bugcov-ipc}) concentrate in high-privilege
processes, yet academic browser fuzzers are largely absent. Only
SbxBrk~\cite{bars2025empirical} explores sandbox-escape testing,
by fuzzing the IPC channel from the V8 sandbox into the
JavaScript engine.
Both surfaces are otherwise reached only indirectly through
whole-browser fuzzing.

Projecting these fuzzers onto the bug-density map of
\autoref{sec:bugcov} answers RQ3. Academic works cluster on
the script and document inputs: twenty JavaScript/Wasm and seven
document fuzzers, against one shader fuzzer, one IPC fuzzer, and
none for UI gestures. Hence, the UI and IPC surfaces lack
dedicated fuzzers, while the WebAPI bindings, the PDFium SDK
boundary, and the WebGL backend have applicable techniques, but
deployment gaps keep their coverage low.

\subsection{Overlooked Deployment Gaps}
\label{ssec:fuzzer-missing}

Most techniques in~\autoref{ssec:fuzzer-axes} focus on the
Mutation (\textbf{M}), Generation (\textbf{G}), Intermediate
Feedback (\textbf{F}), and Bug Oracle (\textbf{O}) stages.
Researchers inherit the vendor deployment and assume it is correct.
However, \autoref{sec:bugcov} suggests otherwise: the deployment
itself leaves gaps. We observe three recurring gaps.

\textbf{Gap 1: incorrect configuration.}
Document fuzzers~\cite{domato,dinh2021favocado,xu2020freedom} rely on a
complete list of the target API interfaces to cover the corresponding
components. Even though the required grammars are well defined in the
codebase, a harness can still fail to import them, leaving whole modules
untested. For instance, in the Chromium tree, the Domato
engine~\cite{domato} ships well-defined grammars (\eg WebGL, WebGPU),
yet the \texttt{domato\_html\_in\_process\_fuzzer} harness that drives
Domato does not import them. As a result, the corresponding WebAPI
modules are barely exercised: \texttt{modules/webgl} at 0.1\%
and \texttt{modules/webgpu} at 3.6\% (\autoref{ssec:bugcov-doc}). These are
not algorithmic gaps but developer configuration mistakes, and they can
be resolved by importing the existing grammars into the 
fuzzing engine.

\textbf{Gap 2: oversimplified harness.}
Developers write dedicated fuzzing harnesses to directly expose
deeper internal interfaces to the fuzzer. While such hand-written
harnesses follow the API semantics, they are not always
effective~\cite{gorz2025empirical}. In the pursuit of simplicity, a
developer may omit optional callback fields that the target only
invokes when they are supplied. Browsers, by contrast,
leverage those fields, so the omission silently elides the
code paths that the browser exercises. For
instance, we observe that the poor coverage of PDFium's
\texttt{fpdfsdk} layer originates from null-filled embedder
callbacks. Concretely, the shared driver (\lstinline{testing/fuzzers/pdfium_fuzzer_helper.cc}) on which every PDFium fuzzer relies leaves the callback tables null, even though Chromium's production PDF integration populates them. 
Since \texttt{fpdfsdk} guards each callback invocation with a 
non-null check, the form-fill, JavaScript-platform, and named-action dispatch paths are never entered. Similar issues also arise in libpng, 
media and webcodec fuzzers, suggesting the necessity of fixing such issues.

\textbf{Gap 3: missing harness.}
Most academic works fuzz the whole browser or JavaScript engine to stay
faithful to the threat model. However, driving the entire target is slow
and tends to miss attack surfaces that are hard to reach from the
top-level entry point. Dedicated harnesses that expose these components
directly are therefore needed for thorough testing. In practice, writing
such harnesses costs developer effort, so several bug-dense components
still lack any dedicated harness. 
For instance, IPC receivers (\autoref{ssec:bugcov-ipc}) are not
adequately exercised by whole-browser fuzzing alone, leaving
their logic untested despite a large volume of historical bugs.
Building dedicated harnesses for these components is thus a concrete,
one-time investment that recovers the missing coverage.

\section{Discussion}
\label{sec:discuss}

\textbf{Browser testing techniques beyond fuzzing.}
While fuzzing is the most widely deployed automated testing
technique against browsers, vendors and security researchers
apply complementary techniques as well, including manual code
audits~\cite{chrome_code_review}, static analysis~\cite{brown2017finding}, symbolic
execution~\cite{brown2020sys,han2021precise}, and LLM-based
testing~\cite{p0_bigsleep}. Different from fuzzing, these techniques 
are not tied to a specific input class and can in principle reach any
component, directly targeting the deeply nested privileged code, such as
the UI event path or the IPC dispatch handlers. 
However, many of them rely either on human expertise or on 
component-specific heuristics that do not scale to the whole browser.
Manual code audits depend on the reviewer's familiarity with the
target subsystem and cover only the fraction of code. 
Static analysis avoids that bottleneck but
suffers from a higher false positive rate. Symbolic and concolic 
execution reason more precisely about reachable states, 
but path explosion limits their application. 
LLM-based testing is a new emerging direction and we further 
discuss below.

\textbf{Limitations.}
Three limitations may threaten the validity of this systematization. 
The first concerns the precision of our data processing. Most
fields in each vulnerability report are extracted by string
keyword matching against shepherd-labelled metadata; a small
number are derived from LLM-based parsing of the report text.
\autoref{sec:bugcov} describes the pipeline and
the manual validation we ran on a sample of \numSampleReport reports.
The second concerns the interpretation of coverage values. 
There is no fixed line that separates well-tested 
code from under-tested code, and the right cut-off can differ across
components. We therefore use 30\% as a working threshold in
this paper. We do not claim this value is precise, but it shows 
the relative trend across components, which is
how our findings should be interpreterd.
The third concerns data availability. Our systematization is
limited to the data that vendors release publicly. In
particular, Safari does not publicly release bug
reports, so we cannot include it in the dataset. To avoid
overfitting our conclusions to a single vendor, we collect both
Chrome and Firefox reports and compare their counts and
distributions.

\textbf{LLM-based browser vulnerability discovery.}
The improving capability of large language models is now
enabling the detection of deeper bugs in real-world software.
Early attempts, limited by the model capability, only discovered 
vulnerabilities in smaller-scale projects such as SQLite~\cite{p0_bigsleep}. However, recent frontier models have reported vulnerabilities in the JavaScript engine and the browser itself~\cite{anthropic_firefox}. 
This shift suggests a fast improvement in capability and points to
substantial potential for future LLM-based applications.
Compared with traditional fuzzing, LLM-driven testing can
generate diverse inputs and resolve complex API
dependencies, which allows deeper exploration of the attack surface,
though typically at a higher cost per test. 
These properties make LLM-driven testing a promising direction for the
overlooked testing surfaces identified in \autoref{ssec:fuzzer-missing}, where existing public efforts are insufficient.

\section{Lessons From the Systematization}
\label{sec:lesson}

\textbf{Principle of least privilege is violated.}
By design, a web browser should follow the principle of least
privilege (PoLP): untrusted (web) content must not be handled
in privileged code~\cite{chrome_least_privileg}.
However, \autoref{sec:threat-model} suggests this principle
is not always applied.
For instance, the moderate-privilege GPU process takes graphic
shaders (attacker-controlled input) directly,
allowing the attacker to corrupt privileged code.
The same pattern affects the network process (network packets)
and the Firefox browser process (certificate validation).
An ideal browser should enforce the PoLP to
isolate and minimize the risks.

\textbf{Testing investment is misaligned with bug yield.}
Despite extensive fuzzing efforts from industry and academia,
the coverage profile of \autoref{sec:bugcov} shows that
several components remain systematically under-tested precisely
where they contribute the most memory corruption bugs. The
pattern recurs across input classes: bug-heavy surfaces such as
Blink's WebAPI bindings (\eg WebGL, WebGPU), PDFium's
SDK boundary (\texttt{fpdfsdk}), the UI input path
(\emph{BrowserChrome}, \emph{DevToolsWebUI}), and the IPC
dispatch targets (\texttt{chrome/browser/}, \texttt{content/browser/}) are 
all under-covered. The structural cause is that testing 
concentrates where mature techniques exist 
(\eg binary CGF, V8 engine fuzzing) and largely skips 
surfaces for which no public harness was ever written.

\textbf{Beyond the algorithm: fix the deployment mistakes.}
The fuzzing literature surveyed in \autoref{sec:browser-testing}
optimises four algorithmic stages under the
implicit assumption that a developer correctly deploys the
fuzzer at every reachable surface. The coverage 
gaps in \autoref{sec:bugcov} suggest this assumption breaks
in three patterns. \emph{Incorrect configuration:} the
algorithm and the input specification both exist, but the 
deployed configuration fails to link them (\eg Blink's WebAPI
IDLs are not imported by Domato's default grammar).
\emph{Oversimplified testing harness:} a harness exists for the
target surface, but the surrounding environment is stubbed too
aggressively (\eg PDFium's reference fuzzer leaves every
form-fill FFI callback null. 
Mojo grammar fuzzers analogously fix optional message fields to
defaults). \emph{Missing harness:} a bug-dense surface has no
dedicated harness at all (\eg the UI input path, where few 
public Chromium fuzzers are available). Each requires a deployment fix 
that no algorithmic improvement can substitute for.

\section{Conclusion}
\label{sec:conclusion}

Despite a decade of intensive browser testing, the community
lacks a unified view of the browser's low-level attack
surface and of how far that testing reaches into it.
This paper systematizes the browser's low-level attack surface
along an Input~$\times$~Component~$\times$~Privilege taxonomy
abstracted from the browser design documents.
We map \datasetMC memory corruption reports
disclosed between 2016 and 2025 onto this taxonomy, pair
each component with the line coverage that vendor-deployed
fuzzers achieve, and overlay a decade of academic browser
fuzzers on the same map. The combined view exposes five
surfaces where testing investment diverges from bug yield
(the WebAPI bindings, the PDFium SDK boundary, the WebGL backend,
the UI input path, and the IPC receivers) and three
recurring deployment gaps (incorrect configuration,
oversimplified harnesses, and missing harnesses).

These findings point to three directions for future browser
security work: finer-grained least-privilege separation that
converts high-privilege bugs into low-privilege ones;
testing-coverage expansion for the five under-tested surfaces;
and closing the three deployment gaps. We hope future work can 
use these findings as a map of the architecture, a measurement
of past effort, and a prioritisation guide for where new effort
is most needed.

\bibliographystyle{IEEEtran}
\bibliography{paper}

\appendix

\section{Appendix}

\subsection{Ethics Considerations}
\label{sec:ethic}

This paper studies disclosed vulnerabilities in the browser
ecosystem. While we point out future directions for
vulnerability research in the browser, the paper itself does
not disclose any new vulnerability and therefore does not
threaten the browser system. Moreover, we collect only
public reports from vendor issue trackers. The entire data
collection and processing pipeline involves no work with
live systems or human-subject studies. We therefore consider
that this work does not involve any ethics considerations.

\subsection{Paper Selection Criteria}
\label{sec:append-criteria}

We collect papers published at top-tier security and
software-engineering venues, including IEEE S\&P, USENIX
Security, NDSS, ACM CCS, ICSE, ISSTA, ASE, and FSE.

For documents, we search the keywords ``DOM'', ``Browser'', and
``Binding'' to find HTML testing papers, and further search
``PDF'' to identify PDF engine testing papers. We exclude
server-side bug-finding works, such as those targeting SSRF and
XSS detection~\cite{drescher2025dom,liu2025domino}, and works
that target only functional bugs~\cite{song2022r2z2,song2023metamong}.

For scripts, we search the keywords ``JavaScript'',
``WebAssembly'', and ``Wasm''. We exclude papers that target
web applications or Node.js 
packages~\cite{xiao2024jasmine,kang2023scaling}, keeping only
works that target vulnerabilities in JavaScript or Wasm engines.

For graphics, we search the keywords ``WebGPU'' and ``WebGL'',
the two major graphics APIs exposed by modern browsers.

For UI and IPC, we search the keywords ``GUI'' and ``IPC''. We
find no academic works that specifically target browser UI or
IPC vulnerabilities.

For binary inputs, we search the keyword ``Fuzz'' to cover
general greybox fuzzing techniques.

\end{document}